\documentclass[11pt]{article}
\usepackage{
amsmath,amsfonts,amssymb,amsthm,bm,calrsfs,stmaryrd}

\usepackage{times}

\def\bea{\begin{eqnarray}}
\def\eea{\end{eqnarray}}
\def\be{\begin{equation}}
\def\ee{\end{equation}}
\def\ba{\begin{array}}
\def\ea{\end{array}}
\def\eq#1{(\ref{#1})}
\def\yb{\bar{y}}
\def\zb{\bar{z}}

\def\kappab{\bar{\kappa}}

\def\cI{\mathcal{I}}

\def\bea{\begin{eqnarray}}
\def\eea{\end{eqnarray}}
\def\ba{\begin{array}}
\def\ea{\end{array}}

\def\a{\alpha}
\def\b{\beta}

\def\d{\delta}
\def\e{\epsilon}
\def\r{\rho}
\def\s{\sigma}
\def\t{\tau}
\def\r{\rho}
\def\C{\Gamma}

\def\L{\Lambda}
\def\O{\Omega}
\def\o{\omega}

\def\Th{\Theta}
\def\ad{\dot{\alpha}}

\def\una{{\underline{\alpha}}}
\def\unb{{\underline{\beta}}}

\def\ad{{\dot{\alpha}}}

\def\unL{{\underline{\Lambda}}}
\def\l{\lambda}
\def\lb{\bar{\lambda}}
\def\m{\mu}
\def\mb{\bar{\mu}}

\def\Thb{\bar{\Theta}}

\def\charti{I}

\def\Real{\mathbb{R}}
\def\Comp{\mathbb{C}}

\def\U{\widehat{U}}
\def\V{\widehat{V}}
\def\F{\widehat{\Phi}}
\def\J{\widehat{J}}
\def\A{\widehat{A}}

\def\msp{\mathfrak{sp}}
\def\mso{\mathfrak{so}}

\def\mhs{\mathfrak{hs}}

\def\NumbCurv{{K}}

\def\ft#1#2{{\textstyle{{\scriptstyle #1}
\over {\scriptstyle #2}}}}

\usepackage{color}
\definecolor{rougef}{rgb}{0.56,0,0}
\definecolor{vertf}{rgb}{0,0.5,0}
\definecolor{bleuf}{rgb}{0,0,0.8}

\begin{document}

\begin{center}
{\LARGE\bf  Twistor Space Observables and Quasi-Amplitudes \\[10pt] in 4D Higher-Spin Gravity}
\vspace{1cm}

{\Large Nicol\`o Colombo\footnote{Ulysse Incentive Grant for Mobility in Scientific Research, F.R.S.-FNRS ;~ {\texttt{nicolo.colombo@umons.ac.be}}}~ and~~Per Sundell\footnote{Ulysse Incentive Grant for Mobility in Scientific Research, F.R.S.-FNRS ;~ {\texttt{per.sundell@umons.ac.be}}}}

\vspace{1cm}

{Service de M\'ecanique et Gravitation\\[5pt] ~Universit\'e de Mons --- UMONS \\[5pt] 20 Place du Parc, B-7000 Mons, Belgium}\\
\vspace{.5cm}
\end{center}

\vspace{.5cm}
\begin{minipage}{.95\textwidth}
\begin{center}\large{\bf{Abstract}}\end{center}

\quad Vasiliev equations facilitate globally defined formulations of higher-spin gravity in various correspondence spaces associated with different phases of the theory. In the four-dimensional case this induces a correspondence between a generally covariant formulation in spacetime with higher-derivative interactions to a formulation in terms of a deformed symplectic structure on a noncommutative doubled twistor space, whereby spacetime boundary conditions correspond to sectors of an associative star-product algebra. In this paper, we look at observables given by integrals over twistor space defining composite zero-forms in spacetime that do not break any local symmetries and that are closed on shell. They are nonlocal observables that can be evaluated in single coordinate charts in spacetime and interpreted as building blocks for dual amplitudes. To regularize potential divergencies arising in their curvature expansion from integration over twistor space, we propose a closed-contour prescription that respects associativity and hence higher-spin gauge symmetry. Applying this regularization scheme to twistor-space plane waves, we show that there exists a class of dual amplitudes given by supertraces. In particular, we examine next-to-leading corrections, and find cancellations that we interpret using transgression properties in twistor space.

\end{minipage}

\newpage

{\small \tableofcontents }

\section{Introduction}

\subsection{Summary of our results}

In this paper, we examine a particular type of classical observables of Vasiliev's four-dimensional bosonic higher-spin gravities \cite{Vasiliev:1990en,Vasiliev:1990cm} (see also \cite{Vasiliev:1990vu,Vasiliev:1992av}),  namely the zero-form charges introduced in \cite{Sezgin:2005pv} and evaluated on exact solutions in \cite{Sezgin:2005pv,Iazeolla:2007wt,Iazeolla:2008ix}; see also \cite{Boulanger:2008up} and the forthcoming papers \cite{Sezgin:2011hq,WipCarlo}. 
These observables are defined in terms of the full master fields of the theory.
They can be expanded in terms of non-local functionals of the dynamical scalar field and the on-shell curvatures of the dynamical gauge fields with positive integer spin, known as generalized Weyl tensors.
The aim of our paper is to examine their curvature expansion in more detail in certain sectors of the theory.

In unfolded dynamics \cite{Vasiliev:1988xc,Vasiliev:1988sa}, the scalar as well as the generalized Weyl tensors are treated together with all their nontrivial space-time derivatives on the mass shell as independent differential zero-form fields. 
The unfolded formulation thus amasses an infinite-dimensional set of zero-forms, whose integration constants contain the local degrees of freedom of the theory.
This set constitutes a single master field, referred to as the Weyl zero-form,
taking its values in a unitarizable representation of the higher-spin algebra, the twisted-adjoint representation. 
The zero-form charges are thus functionals of the Weyl zero-form that are closed on shell and that hence can be evaluated at some arbitrarily chosen point in spacetime.
Their being on-shell closed is equivalent to that they do not break any higher-spin gauge symmetries, which is important for their physical interpretation\footnote{In particular, it means that they can be used to enrich the parity-violating interaction ambiguity of bosonic higher-spin gravities \cite{Sezgin:2011hq}.}.

A key feature of Vasiliev's formulation of four-dimensional higher-spin gravity is that the twisted-adjoint representation space, where thus Weyl zero-form belongs, has a dual description in terms of functions on the two-complex-dimensional twistor space treated as a noncommutative manifold with a star product. 
In this fashion, twistor-space boundary conditions correspond to boundary conditions on the dynamical fields in spacetime.
The higher-spin symmetry then organizes various sets of such duals pairs of boundary conditions into  irreducible representations that make up different sectors of the theory. This leads to a key physical problem, namely to determine which combinations of such sectors constitute globally well-defined formulations of four-dimensional higher-spin gravity.

In this paper, we shall focus on a particular sector of the theory, consisting of twistor-space plane waves, as originally introduced in \cite{Vasiliev:1990bu,Bolotin:1999fa}. 
The perturbative completion of classical solutions starting from linearized twistor space plane waves,  has been studied in \cite{Vasiliev:1990vu,Prokushkin:1998bq}, where in particular the issue of convergence of the star-products was analyzed, as we shall review briefly in what follows.
As was established in \cite{Vasiliev:1990vu,Prokushkin:1998bq}, the plane-wave sector is free of any divergencies at the level of the locally defined fundamental master fields.


In what follows, we shall examine the effect of inserting these perturbatively defined solutions into the classical observables given by the zero-form charges introduced in \cite{Engquist:2005yt,Sezgin:2005pv}\footnote{For further constructions of zero-form charges, see \cite{Sezgin:2005hf,Iazeolla:2007wt}. The r\^oles of massive parameters and the related notion of a dual Weyl zero-form in strictly massless theories are addressed in \cite{Boulanger:2008up}. For similar constructions in the case of de Sitter gravity,  see \cite{Giddings:2007nu}.}. A key feature of the zero-form charges are that they are functionals of the Weyl zero-form, which is a locally defined object, that are invariant under all higher-spin gauge symmetries off shell. This implies that the zero-form charges are globally defined off shell and de Rham closed on shell. Unlike generic intrinsically defined and hence nonlocal observables, which typically requires many space-time charts for their evaluation, the zero-form charges can be evaluated in a single chart and interpreted as basic building blocks for dual twistor-space amplitudes, as we shall discuss below.
The zero-form charges are given by traces of star-products of the Weyl zero-form. The trace operations is defined by integration over twistor space with insertions of various combinations of holomorphic and anti-holomorphic Klein operators, realized using Weyl ordering as Dirac delta functions. While the star-products have finite curvature expansions in the plane-wave sector, in accordance with \cite{Vasiliev:1990vu,Prokushkin:1998bq}, potential divergencies arise in the trace operation. 

In order to regularize these, we propose a perturbatively defined prescription whereby the auxiliary open-contour integrals appearing in the curvature expansion of the master fields \cite{Vasiliev:1990en,Vasiliev:1990cm} are replaced by auxiliary closed-contour integrals via an insertion of a logarithmic branch cut; see Eq. \ref{j integration}. As we shall show in Section \ref{Large-contour prescription}, in the sector of twistor-space plane waves, this yields a well-defined curvature expansion of the particular type of zero-form charges obtained by inserting both types of Klein operators into the trace. In this case, the potential divergencies can be avoided by deforming the closed contours while preserving associativity and hence higher-spin gauge symmetry. 
Working within this scheme, we examine next-to-leading corrections and find cancellations that we interpret using transgression properties in twistor space.

We also encounter other formally defined gauge-invariant functionals that have actual divergencies in the sector of twistor-space plane waves. These objects may have two interpretations. One is that they are simply ill-defined as physical observables and should not be considered at all. An alternative approach, is to instead examine whether they can be regularized in other sectors of the theory, such as unitarizable representations of the higher-spin algebra, which we leave for future studies. Our paper contains a number of comments pertaining to the latter more general picture, that by now is starting to become clearer, and that we hope will stimulate further progress in this field.

\subsection{Plan of the paper}
\noindent The rest paper is organized as follows:

The remainder of Section 1 contains further general remarks on how the unfolded approach lends itself naturally to studying observables and semi-classical localizability.

In Section \ref{sec vasiliev system} we review the Vasiliev equations in the case of four-dimensional minimal bosonic higher-spin gravities, and outline their curvature expansion in ``twistor gauge'' which yields a unique perturbative expansion for real-analytic initial data in twistor space. We also discuss the 
unfolded treatment of initial and boundary values and the corresponding notion of moduli spaces of globally defined classical solutions. 

In Section 3 we examine further the key r\^ole played by the Weyl zero-form in unfolded dynamics, and we present locally accessible zero-form observables and discuss how these can be used to examine the localizability of states. 

In Section \ref{Large-contour prescription} we propose a perturbatively defined closed-contour scheme for regularizing potential divergencies in star products and traces in twistor-space that preserves associativity and hence higher-spin gauge invariance and that reduces to the open-contour scheme for sufficiently regular initial data. We then apply this scheme to the curvature expansion of zero-form invariants in the sector of twistor-space plane waves. We find that several observables, based on supertraces, remain uncorrected in the next-to-leading order which we interpret using a transgression formula in twistor space.

Finally, in Section 5 we conclude by summarizing our results and outlining future directions.

In Appendix \ref{normal-ordered symbols} we fix our conventions for the star product.

\subsection{Observables, regularization and localizability}

In generally covariant field theories, the classical solution spaces, or moduli spaces, consist of gauge equivalence classes of boundary conditions.
The classical functions on these spaces, or classical observables,
are intrinsically defined functionals of the locally defined fields, \emph{i.e.} functionals that are gauge invariant off shell and diffeomorphism invariant on the base manifold on shell; for details in the case of higher-spin gravity, see the forthcoming papers \cite{Boulanger:2011dd,Sezgin:2011hq}. 
In this sense,  there is an intimate interplay between the choices of on-shell observables and off-shell structure group.
This leads to the notion of a topological symmetry breaking mechanism that induces various moduli spaces, which one may refer to as homotopy phases of a generally covariant field theory. 

Visualizing these moduli spaces using classical observables, the physically relevant issues are i) the dynamical nature of the topological symmetry breaking mechanism, \emph{i.e.} its implementation at the level of  a path integral (for discussion in the case of higher-spin gravity, see \cite{Boulanger:2011dd,Sezgin:2011hq}); and ii) whether spaces of linearized initial data belonging to unitary, or unitarizable, representations of the gauge algebra can be completed, perturbatively or by exact methods, into spaces of globally defined full solutions forming subspaces of suitable homotopy phases. 
On physical grounds, one expects that unitarizability arises in sectors of the theory consisting of multi-body solutions that factorize in limits into products of single-body solutions occupying finite regions of background spacetimes. 
The factorization should arise from imposing suitable boundary conditions on the base manifold as well as in target space, leading to multi-body systems in which each body has a well-defined center-of-mass.
As their spatial separation becomes large, one may then require that the separate bodies decouple from each other in the sense that the classical observables exhibit cluster decomposition, as we shall discuss in more detail below. 
This form of localizability should be independent of non-localities appearing via gauge artifacts or in higher-derivative interactions in the locally defined effective equations of motion. 

Physically speaking, at scales far from cosmological or Planckian regimes, it makes sense to sidestep and temporarily postpone the study of the aforementioned issues in the case of general base manifolds, and begin by focusing on perturbative expansions of amplitudes around background metrics of simple topology and with boundary conditions imposed on the dynamical fields.
The resulting amplitudes are holographic observables tied to the boundaries \cite{Giddings:2005id}, serving as the basic building blocks for a topological sum, or as generating functions, in suitable limits, of localized bulk observables such as relational observables \cite{Gary:2006mw} and flat-space scattering matrix elements \cite{Gary:2009ae}. 
This leads to the notion of order parameters for soldered and metric phases. These are the holographic observables and other classical observables that depend on metric structures, such as homotopy charges, minimal areas and possibly partition functions of tensile branes. In the off-shell formulation, these order parameters break the gauge symmetries of the soldering one-form, \emph{i.e.} the local translations.
\footnote{Off shell, the soldering one-form belongs to a section of the gauge bundle associated to the principle gauge bundle of the structure group \emph{i.e.} the group generated by the unbroken gauge parameters; for further details in the context of higher-spin gravity, see the forthcoming paper \cite{Sezgin:2011hq}. In the metric phase, the soldering one-form is assumed to be invertible. On shell, its gauge parameters belong to a section of the gauge bundle and are associated to globally defined vector fields thus identifiable with diffeomorphisms.}

There are also observables that do not break any local symmetries, including locally defined translations, and which hence remain valid in the unbroken phase.
A particular class of such observables are locally accessible in the sense that they can be evaluated using the field content of a single coordinate chart. In the unfolded formulation, these observables are composite zero-forms ${\cal I}={\cal I}(\Phi)$ where $\Phi$ is the Weyl zero-form \cite{Vasiliev:1988xc,Vasiliev:1988sa}; see also below. These observables are valid also in the soldered and the metric phases, where they can be expanded in terms of all possible on-shell derivatives of the locally defined fields.
Unlike the order parameters, these observables remain nontrivial as the metric background degenerates or becomes insignificant in comparison to the size of metric fluctuations. Their not breaking any gauge symmetries, \emph{i.e.} $\delta_\e{\cal I}=0$, is equivalent to that they are globally defined off shell and de Rham-closed on shell, \emph{i.e.} $d{\cal I}=0$ modulo the equations of motion.
One may hence refer to them as zero-form charges.

There are several ways of interpreting the zero-form charges. In the classical theory, they have a natural interpretation as Casimir invariants for locally defined systems of linearized unfolded equations of motion \cite{Boulanger:2008up}.
Two such systems, associated to two overlapping coordinate charts, can be glued together on shell only if all zero-form charges agree. In the quantum theory, one may instead seek to interpret them as basic building blocks for amplitudes, namely the on-shell values of certain deformations \cite{Sezgin:2011hq} of the topological action principle of \cite{Boulanger:2011dd} generalizing the action principle of \cite{Vasiliev:1988sa} recently revisited in \cite{Doroud:2011xs}.  In this context, their perturbative $\Phi$-expansions, \emph{viz.} ${\cal I}(\Phi)=\sum_{n=0}^\infty {\cal I}^{(n)}(\Phi,\dots,\Phi)$, yield multi-linear and bose-symmetric functionals ${\cal I}^{\,(n)}(\Phi_1,\dots,\Phi_n)$, that we refer to as quasi-amplitudes. One may ask:
\begin{itemize}\label{page:twokey}
\item {\bf How to regularize quasi-amplitudes}?\\
Let us precise the question as follows: The candidate unitarizable representations of the higher-spin algebra arise in the abstract Weyl zero-form module as the result of choosing boundary conditions. In the presence of a finite cosmological constant, these representations have the property of being isomorphic to their duals; for a general discussion, see for example \cite{Boulanger:2008up,Boulanger:2008kw}
\footnote{For strictly massless models, such as Yang-Mills theory in flat 
spacetime and gravity with vanishing cosmological constant, the construction of 
locally accessible observables appears to require an extension of 
the Weyl zero-form by a dual Weyl zero-form $\Phi^\ast$ 
containing unfolded generalizations of vacuum expectation values 
\cite{Boulanger:2008up,Boulanger:2008kw}.}.
More generally, direct products of such self-dual representations may contain singlets, which correspond to zero-form charges of the free theory arising in the leading order of the $\Phi$-expansion.
One may then examine whether these free-theory zero-form charges can be dressed by sub-leading equivariant corrections into perturbatively defined zero-form charges of the full theory.
In the metric phase, these corrections are expansions in derivatives of fluctuations of the dynamical fields, given in units of the cosmological mass-scale. 
In unitarizable sectors, these expansions, which are now taken on shell with specific boundary conditions, may become strongly coupled in which case their evaluations require regularization schemes, that may be specific to the sectors under study.

\item \label{pagecluster} {\bf How to identify sectors of localizable states}?\\
Let us precise the question as follows: Independently of whether the equations of motion contain nonlocal interactions or not, 
the physically relevant question is whether the theory admits boundary conditions corresponding to unitarizable sectors of states $\{f^{s_i}_{p_i}\}$ that are i) labeled by points $p_i$ and internal labels $s_i$; and ii) localizable in spacetime in the sense that the quasi-amplitudes  $ {\cal I}^{~(n)}(f^{s_1}_{p_1},\dots,f^{s_n}_{p_n})$ fall off sufficiently fast as the points $p_i$, $i=1,\dots,n$, are separated spatially;       \emph{c.f.} the fall-off behaviors of holographic amplitudes \cite{Giombi:2009wh,Giombi:2010vg} and the curvature tensors in the one-body soliton solutions of
\cite{Didenko:2006zd,Didenko:2008va,Didenko:2009tc,Didenko:2009td,WipCarlo}.
\end{itemize}

In what follows, we shall examine these two issues in mode detail in the case of Vasiliev's four-dimensional higher-spin gravity where the zero-form charges are given by integrals over a twistor space \cite{Engquist:2005yt,Sezgin:2005pv,Sezgin:2005hf,Iazeolla:2007wt}.

\subsection{Generalities of unfolded dynamics}

To examine whether metric phases can be generated dynamically by perturbing unbroken phases by metric order parameters, it is natural to start from Vasiliev's unfolded dynamics \cite{Vasiliev:1988xc,Vasiliev:1988sa}. The reason is that unfolded dynamics provides manifestly diffeomorphism-invariant parent formulations of generally covariant quantum field theories in which i) the locally defined classical field dynamics is described by a topological field theory that does not refer to a non-degenerate metric background; ii) effective frame-like formulations arise upon perturbative eliminations of auxiliary fields assuming that a soldering one-form is invertible (whether or not the graviton is dynamical); and iii) the transition between topological and metric phases is smooth at the level of counting locally accessible and gauge invariant degrees of freedom, as we shall discuss in more detail below. 
In other words, these unfolded parent formulations disentangle the two r\^oles usually played by the metric as gauge field for local translations as well as carrier of local spin-two degrees of freedom.
As a result, the former r\^ole arises upon soldering while the latter r\^ole is taken over by an independent spin-two Weyl zero-form field\footnote{In particular, ordinary relativistic quantum field theories in rigid metric backgrounds, such as flat spacetime, arise as spontaneously broken phases of diffeomorphism invariant topological field theories with dynamical vielbein and Lorentz connection and fixed, non-dynamical spin-two Weyl zero-form, vanishing for flat spacetime.}

The aforementioned features are innate in unfolded dynamics since it is based on the formulation of field theory starting from algebraic structures that are more rudimentary than metric structures, namely various graded differential algebras, ranging from the free and graded commutative case to the strongly homotopy associative one, via the quasi-free and associative graded differential algebras, which are the ones of relevance for Vasiliev's higher-spin gravities.
The graded commutative case, \emph{i.e.} which is the natural first differential-form generalization of the theory of fiber bundles associated to principal bundles for Lie groups over classical base manifolds, was explored already in the pioneering works of Cartan and other early mathematicians (for example, see \cite{Bryant:1991zz} for a review). It was then refined by Sullivan \cite{Sullivan1}, and furthermore brought in contact with supergravities by \cite{Castellani:1981um,D'Auria:1982nx,vanNieuwenhuizen:1982zf,Castellani:1982ke,D'Auria:1982pm} though in a hybrid set-up that exploits only partially the utilities of differential algebras. 

These were taken into account more fully by Vasiliev \cite{Vasiliev:1988xc,Vasiliev:1988sa} in the context of reconciling higher-spin and general covariance on shell\footnote{For reviews on higher-spin gauge theories, see \cite{Vasiliev:1995dn,Vasiliev:2000rn,Bekaert:2005vh}; \cite{Bengtsson:2008mw} which stresses formal structures and third-quantization; and \cite{Bekaert:2010hw} which is a non-technical review of the key mechanisms going into the higher-spin extensions of ordinary gravity.}. 
In doing so, Vasiliev identified the key r\^oles played by i) infinite-dimensional Weyl zero-form modules in deforming gauge structures on shell \cite{Vasiliev:1988xc,Vasiliev:1988sa}; and ii) the natural generalization of free and graded commutative differential algebras to quasi-free and associative dittos on noncommutative base manifolds and correspondence spaces. These two refinements together form the cornerstones in his monumental works 
\cite{Vasiliev:1990en,Vasiliev:1990cm} (see also \cite{Vasiliev:1990vu,Vasiliev:1992av}) on fully nonlinear unfolded equations of motion for four-dimensional higher-spin gravities\footnote{See also \cite{Iazeolla:2007wt} for generalizations to various signatures including chiral models in Kleinian and Euclidean signatures.}, later extended to lower dimensions \cite{Vasiliev:1992ix,Vasiliev:1995sv}\footnote{See also \cite{Vasiliev:1997dq,Prokushkin:1998bq} containing an interesting mechanism of 
relevance to topologically massive and/or chiral gravities in three dimensions. 
We note that three-dimensional higher-spin gravities without matter 
\cite{Blencowe:1988gj} sit on-shell as consistent truncations of corresponding 
matter-coupled Vasiliev systems obtained by setting all zero-forms to 
zero.} 
as well as symmetric tensor gauge fields in higher dimensions 
\cite{Vasiliev:2003ev}\footnote{See also \cite{Sagnotti:2005ns} for an 
alternative trace-unconstrained formulation.}. 

We remark that the generalization of unfolded dynamics to strongly homotopy associative graded differential algebras is based on generalized Hamiltonian quantum field theories in more than one dimension. These theories have been developed, largely independently of unfolded dynamics, within topological AKSZ-BV field theory \cite{Alexandrov:1995kv,Hofman:2000ce,Park:2000au,Hofman:2001zt,Ikeda:2001fq,Hofman:2002rv,Hofman:2002jz} and later adapted to Vasiliev's correspondence-space formalism in \cite{Barnich:2005ru,Barnich:2010sw}. Within this context, it is natural to impose a quantum version of Weyl's Gauge Principle, whereby Nature is to be described by hierarchic duality web consisting of unfolded quantum field theories. As one goes upward in the hierarchy, ghost number at one level becomes identified with form degree at the next level in such a way that the master equation and topological summation uplifts to unfolded equations of motion and radiative corrections, respectively. 

In this context, the Vasiliev systems in various dimensions and with different amounts of supersymmetry and other internal quantum numbers have been proposed to be the master theories for i) free (super)conformal field theories restricted to bilinear composites \cite{Sundborg:2000wp,Sezgin:2002rt,Klebanov:2002ja,Giombi:2009wh,Giombi:2010vg} with double-trace sewing operations \cite{Sezgin:2002rt}; and ii) topological open strings in (super)singleton phase spaces \cite{Engquist:2005yt}. This massless duality web has furthermore been proposed to fit into tensionless limits of string and M theories with cosmological constants \cite{Sundborg:2000wp,Sezgin:2002rt,Bianchi:2003wx,Engquist:2005yt}. The Vasiliev systems are then viewed as classically consistent truncations of hitherto unknown massively extended higher-spin gauge theories that are to be the master theories for i) free (super)conformal field theories restricted to multi-linear composites \cite{Sundborg:2000wp,Sezgin:2002rt,Bianchi:2003wx} with double-trace sewing operations \cite{Sezgin:2002rt}; and ii) topological Wess--Zumino--Witten models with spectral flow \cite{Engquist:2005yt,Engquist:2007pr}, critical W-gaugings \cite{Engquist:2005yt} and compatible gaugings corresponding to Vasiliev's deformed oscillator algebra (arising in continuum limits of the topological open strings \cite{Engquist:2005yt}).

Having made these general remarks on what one may arguably refer to as the salient features of unfolded dynamics and the crucial role it has played so far in developing higher-spin gravity, we now turn to the main part of the paper.

\section{Vasiliev's four-dimensional minimal-bosonic higher-spin gravity }
\label{sec vasiliev system}

In this section we present Vasiliev's unfolded formulation of four-dimensional higher-spin gauge theories, including gravity, in the case of bosonic models. The unfolded equations of motion provide a fully nonlinear and background independent description of classical higher-spin gravities\footnote{For a maximally duality extended off-shell description of four-dimensional bosonic models, see \cite{Boulanger:2011dd}.}. 
Their expansions around various backgrounds yield perturbative formulations in terms of different sets of dynamical fields. In particular, in the case of the minimal bosonic models, there exists such a perturbative expansion in terms of a dynamical scalar, a metric and a tower of symmetric tensor-gauge fields, also known as Fronsdal tensors, of even ranks, living on a four-dimensional manifold. In this perturbative formulation, which is along the lines of the Fronsdal Programme \cite{Bekaert:2010hw} and lends itself to physical interpretations in terms of ordinary relativistic field theory, the four-dimensional diffeomorphism invariance is manifest while the higher-spin gauge symmetries are not, and instead hold only formally in a double perturbative expansion in terms of weak fields and derivatives, given in units of a cosmological constant. 
The background independent formulation is crucial, however, for the purpose of providing the theory with a globally defined geometric formulation and related classical observables \cite{Sezgin:2011hq}, which can then of course be expanded perturbatively. 
 
\subsection{Locally defined unfolded equations}

\subsubsection{Unfolded formulation in correspondence space}\label{Sec:2.1}

The unfolded equations of motion amount to constraints on the generalized curvatures of two master differential forms, $\widehat \Phi$ and $\widehat A$, of degrees zero and one, respectively. These fields are elements in the unital, graded noncommutative and associative $\star$-product algebra $\O({\cal C})$ consisting of differential forms on ${\cal C}$,  a noncommutative correspondence space given locally by the product
\be {\cal C}~\stackrel{\rm loc}{\cong}~{\cal B}\times {\cal Y}\ ,\ee
where ${\cal B}$ and ${\cal Y}$ are noncommutative base and fiber manifolds, respectively. Letting $\widehat d$ denote the exterior derivative on ${\cal C}$, and ${\cal V}_{\cal Y}$ the symplectic volume form on ${\cal Y}$,  the generalized curvature constraints are of the form 
\be {\cal V}_{\cal Y}\star\left(\widehat d\widehat\Phi+{\cal Q}^{\widehat \Phi}\right)~=~0\ ,\qquad {\cal V}_{\cal Y}\star\left(\widehat d \widehat A+{\cal Q}^{\widehat A}\right)~=~0\ ,\label{unfoldedsystem}\ee
where $({\cal Q}^{\widehat\Phi},{\cal Q}^{\widehat A})$ are nonlinear structure functions built from exterior $\star$-products of $\widehat \Phi$, $\widehat A$ and closed and central elements on ${\cal C}$ in such a way that the constraints are Cartan integrable, \emph{i.e.} compatible with $\widehat d^2\equiv 0$.

In the case of ordinary differentiable base manifolds, Cartan integrable systems are known as free or quasi-free graded differential algebras depending on whether their integrability holds without any extra algebraic constraints on the basic differential form variables or not, respectively. One may thus refer to Cartan integrable systems on noncommutative and associative base manifolds containing closed and central elements, such as Vailiev's equations, as quasi-free associative graded differential algebras. Although there is no consensus in the literature, we prefer to reserve the term unfolded dynamics for the formulation of field theories using differential algebras in general\footnote{For example, more generally, one may consider unfolded systems based on quasi-free strongly homotopy associative graded differential algebras.}. 

The Cartan integrability of an unfolded system implies that its \emph{locally accessible} degrees of freedom are encoded into the initial data $C$ for the \emph{zero-forms} in the system \cite{Vasiliev:1988xc,Vasiliev:1988sa}; for the relation to harmonic analysis, see \cite{Iazeolla:2008ix,Boulanger:2008up,Boulanger:2008kw}.
Given such an initial data, and assuming boundary conditions on the variables of positive form degree, a solution space to an unfolded system can be constructed from a family of gauge functions; for the role of gauge functions in the context of imposing boundary conditions at $\partial {\cal B}$, see \cite{Boulanger:2011dd,Sezgin:2011hq}.  
In the case of \eq{unfoldedsystem}  \cite{Vasiliev:1990bu}, one may thus construct solution spaces using a gauge function $\widehat L$ and an initial condition
\be C~=~\left.\widehat \Phi\right|_{p_0\times {\cal Y}}\ ,\label{initialdata}\ee
at a point $p_0\in  {\cal B}$; for applications to exact solutions, see \cite{Sezgin:2005pv,Iazeolla:2007wt}, and to amplitude calculations, see \cite{Giombi:2010vg}. Globally defined observables that do not break any higher-spin gauge symmetries, and that are hence independent of $\widehat L$, can the be extracted as functionals of $C$ via zero-form charges \cite{Sezgin:2005pv,Iazeolla:2007wt}, which is to become the main topic below.

The aforementioned unfolded formulation of the initial/boundary value problem in generally covariant field theory,  implies that \eq{unfoldedsystem} is equivalent to its reductions to commuting, $n$-dimensional submanifolds ${\cal M}_n\subset {\cal B}$. To this end, one assumes some boundary values for the components of $\widehat A$ that are in the kernel of the pull-back operation to ${\cal M}_n\times {\cal Y}$. Expanding in the initial data
\be \Phi~=~\left.\widehat \Phi\right|_{{\cal M}_n\times {\cal Y}}\ ,\qquad U~=~\left.\widehat A\right|_{{\cal M}_n\times {\cal Y}}\ ,\ee
then yields a perturbatively defined unfolded system on ${\cal M}_n\times {\cal Y}$ described by a free and graded commutative differential algebra of the form
\be d\Phi+{\cal Q}^\Phi~=~0\ ,\qquad dU+{\cal Q}^U~=~0\ ,\ee
where the structure functions $({\cal Q}^\Phi,{\cal Q}^U)$ are linear and bilinear in $U$, respectively, and given by non-polynomial expansions in $\Phi$.
This reduced system can be solved using a reduced gauge function $L$ and the same initial data for the zero-form, \emph{viz.}
\be C~=~\left.\Phi\right|_{p_0\times {\cal Y}}\label{initialdata2}\ee
as in \eq{initialdata}, assuming that $p_0\in{\cal M}_n$. In this fashion, one may reduce the system all the way down to a four-manifold ${\cal M}_4$; assuming that $U$ contains a non-degenerate vierbein then yields a manifestly generally covariant metric formulation of four-dimensional higher-spin gravity with higher-derivative interactions; for further details, see for example \cite{Sezgin:2002ru}\footnote{For the analogous application to the superspace formulation of four-dimensional higher-spin supergravities, see \cite{Engquist:2002gy}.}.   

The manifest general covariance is a consequence of the fact that the unfolded equations of motion possess manifest Lorentz covariance \cite{Vasiliev:1999ba}; for the explicitly Lorentz-covariantized unfolded equations of motion, see also \cite{Sezgin:2002ru} and \cite{Sezgin:2011hq}. Technically speaking, the Lorentz covariantization is achieved via a field redefinition $U=W+K$ where $K$ is linear in the canonical Lorentz connection and $W$ consists of canonical Lorentz tensors. Since the zero-form sector on ${\cal M}_n$ is unaffected by these steps, we shall work mainly with the fields $U$ and its uplift $\widehat U$ on ${\cal C}$.  

\subsubsection{Minimal-bosonic master fields}

In the case of the four-dimensional minimal-bosonic models based on the minimal higher-spin Lie algebra $\mhs(4)\supset \mso(2,3)$, the base manifold
\be {\cal B}~\stackrel{\rm loc}{\cong}~  T^\ast{\cal M}\times {\cal Z}\ ,\ee
that is, the correspondence space ${\cal C}\stackrel{\rm loc}{\cong}T^\ast{\cal M}\times {\cal Z}\times {\cal Y}$, where $T^\ast{\cal M}$ is a phase space with canonical coordinates $(X^M,P_M)$, and ${\cal Z}$ and ${\cal Y}$ are two copies of the complex two-dimensional twistor space with globally defined canonical coordinates $Z^\una=(z^\a,-\zb^\ad)$ and $Y^\una=(y^\a,\yb^\ad)$ forming $\msp(4)$-quartets splitting into $\mathfrak{sl}(2;\Comp)$ doublets\footnote{We use the conventions $\Lambda^{\una}=C^{\underline{\a\b}}\Lambda_{\unb}$ and $\l^\a=\e^{\a\b}\l_\b$ and $\l_\a=\l^\b\e_{\b\a}$ for $\L^{\una}=(\l^\a,\pm \bar\l^{\ad})$, and the notation $\Lambda\cdot \Lambda'=\Lambda^\una \Lambda_\una$, $\l\cdot \l'=\l^\a \l'_\a$ and $\bar\l\cdot \bar\l'=\bar\l^\ad \bar\l'_\ad$\,.}.
The non-vanishing $\star$-commutators  are 
\be[X^M,P_N]~=~i\d^M_N\ ,\quad [Y^{\una},Y^{\unb}]_\star ~=~2i C^{\underline{\a\b}}\ ,\quad [Z^{\una},Z^{\unb}]_\star ~=~-2i C^{\underline{\a\b}}\ ,\label{starXPYZ}\ee
In Eq. \eq{unfoldedsystem}, the sections in ${\cal V}_{\cal Y}\star \O({\cal C})$ are described locally by operators that can in their turn be represented by symbols \footnote{The hats denote quantities that depend generically on both $Y$ and $Z$; we drop the hats in order to indicate reduced quantities that do not depend $Z$\,. }  
$\widehat f(X,P,Z;Y;dX,dP,dZ)$ (see Appendix \ref{normal-ordered symbols}). 
Strictly speaking, to define the theory, the symbols must belong to a space of functions where the $\star$-product rule obeys associativity. The choice of such a space of functions is a key physical problem and one of the key motivations behind the present work. In \cite{Vasiliev:1990vu} (see also \cite{Prokushkin:1998bq}) it has been proposed to work with twistor-space plane waves; these generate a well-defined $\star$-product algebra containing instanton-like exact solutions \cite{Sezgin:2005pv,Iazeolla:2007wt}. Many other applications, however, force the master fields out of this class; for example, see \cite{Iazeolla:2007wt,Iazeolla:2008ix,Didenko:2009td,WipCarlo}.

The exterior derivative on the base manifold ${\cal B}$ reads \be \widehat d~=~d+q\ ,\qquad  d~=~dX^M\partial_M+dP_M \partial^M\ ,\quad q~=~ dZ^\una \partial_\una\ .\label{widehatd}\ee
The duality-unextended master fields of the minimal-bosonic model are a twisted-adjoint zero-form 
\be\widehat \Phi~=~ \widehat \Phi(X,P,Z;Y)\ ,\ee
and an adjoint one-form 
\be \widehat A~=~ \widehat U+\widehat V\ ,\label{hatA}\ee
consisting of a component 
\be \widehat U~=~dX^M \,\widehat U_M(X,P,Z;Y)\,+\,dP_M \,\widehat U^M(X,P,Z;Y)\ ,\ee
along $T^\ast{\cal M}$, and a component  
\be \widehat V~=~dZ^\una\, \widehat V_\una(X,P,Z;Y)\ ,\ee
along the twistor space ${\cal Z}$. In bosonic models, the master fields obey\footnote{Here we are focusing on the models containing spacetimes with Lorentzian signature and negative cosmological constant; for other signatures and signs of the cosmological constant, see \cite{Iazeolla:2007wt}.}
\be \pi\bar\pi(\widehat A,\widehat \Phi)~=~(\widehat A,\widehat \Phi)\ ,\qquad (\widehat A,\widehat \Phi)^\dagger~=~ (-\widehat A,\pi(\widehat \Phi))\ ,\label{bosproj}\ee
where the automorphisms $\pi$ and $\bar\pi$ and the hermitian conjugation are defined by $\widehat d\circ (\pi,\bar\pi,\dagger)=(\pi,\bar\pi,\dagger)\circ \widehat d$ and\footnote{The rule $ (\widehat f\star \widehat g)^\dagger=(-1)^{\widehat f\widehat g}\,\widehat g^\dagger\star \widehat f^\dagger$ holds for both real and chiral integration domain in \eq{starproduct}.}
\bea 
\label{pz1}
\pi~(y_{\a},\yb_\ad;z_\a,\zb_\ad)&=&(-y_{\a},\yb_\ad;-z_\a,\zb_\ad)\ ,\qquad\qquad \pi(\widehat f\star \widehat g)~=~\pi(\widehat f)\star\pi(\widehat g)\ ,\\[5pt]
\bar\pi~(y_{\a},\yb_\ad;z_\a,\zb_\ad)&=&(y_{\a},-\yb_\ad;z_\a,-\zb_\ad)\ ,\qquad\qquad \bar\pi(\widehat f\star \widehat g)~=~\bar\pi(\widehat f)\star\bar\pi(\widehat g)\ ,\\[5pt]
(y_{\a},\yb_\ad;z_\a,\zb_\ad)^\dagger&=&(\yb_{\ad},y_\a;\zb_\ad,z_\a)\ ,\qquad\qquad\quad (\widehat f\star \widehat g)^\dagger~=~(-1)^{\widehat f\widehat g}\, \widehat g^{\,\dagger}\star \widehat f^{\,\dagger}\ .
 \eea
In the minimal bosonic models, the master fields obey the stronger projection condition
\be \tau(\widehat A,\widehat \Phi)~=~(-\widehat A,\pi(\widehat \Phi))\ ,\label{minboscond}\ee
that define define the adjoint and twisted-adjoint representattions of $\widehat{\mathfrak{hs}}(4)$, respectively, and where the anti-automorphism $\tau$ is defined by $\widehat d\circ \tau=\tau\circ \widehat d$ and
\be 
\t~(y_{\a},\yb_\ad;z_\a,\zb_\ad)~=~(iy_{\a},i\yb_\ad;-iz_\a,-i\zb_\ad)\ ,\qquad \t(\widehat f\star \widehat g)~=~(-1)^{\widehat f\widehat g} \t(\widehat g)\star\t(\widehat f)\ .\ee
The $\pi\bar\pi$-projection in \eq{bosproj} and the $\tau$-projection in \eq{minboscond} remove all components of the master fields that are associated with the unfolded description of fermions and symmetric tensors with odd spin, respectively. 

\subsubsection{Master-field equations}

In order to study the zero-form charges, we focus on the models with linear interaction function \cite{Vasiliev:1990en,Vasiliev:1990cm,Vasiliev:1990vu,Vasiliev:1992av} (see also \cite{Sezgin:2002ru,Sezgin:2003pt,Boulanger:2011dd,Sezgin:2011hq}). For this simplest choice, the unfolded equations of motion \eq{unfoldedsystem} amount to that the Yang--Mills-like curvature of $\widehat A$ is equated on shell to the $\star$-product between $\F$ and a deformed symplectic two-form $\widehat J$\,, \emph{viz.}
\be \widehat F+\F\star \widehat J\approx 0\ ,\qquad \widehat F~:=~\widehat d\,\A+\A\star \A\ ,\ee
where $\widehat J$ is defined globally on ${\cal C}$ and obeys \be \widehat d\,\J~=~0\ ,\qquad \left[\J, \widehat f\,\right]_\pi~=~ 0\ ,\qquad \tau(\widehat J\,)~=~\widehat J^{\,\dagger}~=-~\widehat J\ ,\label{Jtwist}\ee 
for any $\widehat f$ obeying
\footnote{The minimal-bosonic model is a consistent truncation of the bosonic model where the $\tau$-projection is replaced by the weaker bosonic projection $\pi\bar\pi(\widehat A,\widehat \Phi)~=~(\widehat A,\widehat \Phi)$\,.} 
$\pi\bar\pi(\widehat f)=\widehat f$ and where we have defined 
\be \left[\widehat f,\widehat g\,\right]_\pi~=~\widehat f\star \widehat g-\widehat g \star\pi(\widehat f\,)\ .\ee
In the minimal model, 
\be \J~=~ -\frac{i}{4} (b \,dz^2~\widehat{\kappa}+\bar b \,d\zb^2~\widehat{\kappab})\ ,\ee 
where $\widehat{\kappa}$ and $\widehat{\bar{\kappa}}$ are the Klein operators\footnote{The two-dimensional complexified Heisenberg algebra 
$[u,v]_\star=1$ has the Klein operator $k=cos_\star(\pi v\star u)$\,, which anti-commutes with $u$ and $v$ and squares to $1$\,. Hence $k$ is invariant under the canonical $SL(2;\mathbb{C})$-symmetry. This property becomes manifest in Weyl order, where the symbol of $k$ is proportional to the two-dimensional Dirac delta function. It follows that $(\kappa,\bar\kappa)$ is invariant under $SL(4;\Comp)\times \overline{SL}(4;\Comp)$\,, that is broken by $dz^2$ and $d\bar z^2$ down to a global $GL(2;\Comp)\times \overline{GL}(2;\Comp)$ symmetry of the Vasiliev system, that is generated by diagonal $SL(2;\Comp)\times \overline{SL}(2;\Comp)$ transformations and the exchange $(y_\a,z_\a)\leftrightarrow (iz_a,-iz_\a)$\,. The latter symmetry is hidden in the formulation in terms of differentials on $Z$-space while it becomes manifest in the deformed-oscillator formulation.}
of the complexified Heisenberg algebra generated by $(y_\a,z_\a)$ and $(\yb_\ad,\zb_\ad)$\,, respectively. The insertion of the inner Kleinians is crucial in order for the deformation $\widehat \Phi\star \widehat J$ of $\widehat F$ on shell to be non-trivial in the sense that it cannot be removed by any field redefinition \cite{Vasiliev:1990en}.

By making use of field redefinitions $\widehat \Phi\rightarrow \l \F$ with $\l\in \Real$\,, $\l\neq 0$\,, the parameter $b$ in $\J$ can be taken to obey
\be |b|~=~1\ ,\qquad {\rm arg}(b)~\in~[0,\pi]\ .\ee  
The phase breaks parity except in the following two cases \cite{Sezgin:2003pt}\footnote{Starting from a general deformation of $\widehat F_{\underline{\a\b}}$\,, one can show that compatibility, manifest Lorentz covariance and unbroken parity lead uniquely to the Type A and Type B models.}:
\be \mbox{Type A model (parity-even physical scalar)}~:~~b=1\ ,\ee 
\be \mbox{Type B model (parity-odd physical scalar)}~:~~b=i\ .\ee 
The integrability of $\widehat F+\F\star \widehat J\approx 0$ implies that $\widehat D(\F\star\widehat J)\approx 0$ with $\widehat D$ acting on $\F\star\widehat J$ in the adjoint representation, that is, 
\be \widehat D\,\F~\approx 0~\ ,\qquad \widehat D\,\F~:=~\F+\A\star\F-\F\star\pi(\A)\ ,\ee
with $\widehat D$ acting on $\F$ in the twisted-adjoint representation.
This constraint on $\widehat \Phi$ is integrable, since $\widehat D^2\F=\widehat F\star \F-\F\star\pi(\widehat F)=-\F\star\widehat J\star \F+\F\star \pi(\F)\star\widehat J= 0$
using the constraint on $\widehat F$ and \eq{Jtwist}.

In summary, fully nonlinear and background independent unfolded formulation of minimal-bosonic higher-spin gravities with restricted (linear) interaction function is given by the generalized curvature constraints
\be \widehat F+\widehat\Phi\star \widehat J~\approx~0\ ,\qquad \widehat D\,\widehat \Phi~\approx~0\ ,\qquad \widehat d\,\widehat J~\equiv~0\  ,\label{VasEoM}\ee
\be \widehat F~:=~\widehat d\,\A+\A\star \A\ ,\qquad \widehat D\,\F~:=~\F+\left[\A,\F\,\right]_\pi\ ,\ee
and the algebraic constraints
\be \left[\widehat A,\widehat J\,\right]_\pi~=~\left[\widehat\Phi,\widehat J\,\right]_\pi=0\ ,\ee
thus forming a quasi-free associative free differential algebra. The consistency of these equations as well as their compatibility with the kinematic conditions \eq{minboscond} is a consequence of the assumed associativity of the $\star$-products among the full master fields $\{\widehat A,\F; \widehat J\,\}$.  

A general property of free differential algebras, is that their Cartan integrability implies Cartan gauge invariance \cite{Vasiliev:1988xc,Vasiliev:1988sa}. In the case of quasi-free associative algebras, the gauge invariance is broken for the central elements (since the gauge transformations must preserve algebraic properties). In the case of \eq{VasEoM}, the Cartan gauge transformations read
\be\delta_{\widehat{\e}} \,\widehat A~= \widehat D \,\widehat \e\ ,\qquad \delta_{\widehat\e}\widehat J~=~0\ ,\ee
\be \delta_{\widehat{\e}} \,\widehat\Phi~=~ -\left[\widehat\e,\F\right]_\pi\ ,\qquad \tau(\widehat\e\,)~=~ -\widehat\e\ ,\qquad (\widehat\e\,)^\dagger ~=~-\widehat\e\ ,\label{gaugetransf}\ee
defining the adjoint and twisted-adjoint representations of the algebra $\widehat{\mathfrak{hs}}(4)$, with closure $[\delta_{\widehat \e_1},\delta_{\widehat \e_2}]=\delta_{\widehat \e_{12}}$ where $\widehat\e_{12}=[\widehat \e_1,\widehat \e_2]_\star$\,.

\subsection{Higher-spin geometries, observables and homotopy phases}
\label{Global formulation: phases and observables}

Before proceeding with the perturbative analysis, we would like to outline the interplay between classical observables and globally defined geometric formulations of higher-spin gravity; for a more detailed presentation, see\cite{Sezgin:2011hq}.
Barring the issue of Lorentz-covariance (see end of Section \ref{Sec:2.1}), the split in \eq{widehatd} and \eq{hatA} yields
\be d\F+\U\star \F-\F\star\pi(\U)~\approx~0\ ,\quad d\U+\U\star \U~\approx~0\ ,\label{unifold}\ee
\be q\U+d\V+\U\star\V+\V\star\U~\approx~0\ ,\label{middle}\ee
\be
\label{e o m}
 q\F+\V\star \F-\F\star\pi(\V)~\approx~0\ ,\quad q\V+\V\star\V + \F\star \J~\approx~0\ ,
\ee
to be given a geometric meaning as a bundle over $T^\ast{\cal M}$. We remark that this system exemplifies a subtlety of quasi-free differential algebras in general, as compared to ordinary free differential algebras: thought of as a bundle-like structures, the topology of the fibers may vary between different regions of the base manifolds. For example, in an exact solution it may be the case that the connection $\widehat V$ may degenerate at some special points or submanifolds of ${\cal M}$ \cite{WipCarlo}.

With this caveat in mind, generalized higher-spin geometries can be defined by treating subalgebras $\widehat{\mathfrak{t}}\subseteq \widehat{\mathfrak{hs}}(4)$ as Lie algebras for structure groups of principal ${\widehat{\mathfrak{t}}}$-bundles over $T^\ast{\cal M}$. In such a geometry, a globally defined solution to the Vasiliev equations \eq{VasEoM} is a gauge equivalence classes consisting of master fields $(\A_\charti,\F_\charti)$ defined on coordinate charts ${\cal M}_I$, labelled 
here by $\charti$\,, glued together by transition functions $\widehat T_
{\charti}^{\charti'}=\exp_\star(\widehat t_{\charti}^{\charti'})$, where $\widehat t_{\charti}^{\charti'}$ are $\widehat{\mathfrak{t}}$-valued functions defined on overlaps. The gauge equivalence relation is given by 
\be \A_\charti ~\sim~  (\widehat G_I)^{-1}\star(\A_I+\widehat d)\star \widehat G_I \ ,\qquad \F_\charti ~\sim~(\widehat G_I)^{-1}\star\F_I\star\pi(\widehat G_I)\ ,\qquad  \widehat T_
{\charti}^{\charti'} ~\sim~  \widehat G^{-1}_\charti\star \widehat T_{\charti}^{\charti'} \star \widehat G_{\charti'}\ ,\ee
where $\widehat G_\charti=\exp_\star(\widehat \e_{\charti})$ with $\widehat \e_{\charti}\in \widehat{\mathfrak{t}}$ defined on ${\cal M}_I$ (without obeying any conditions on $\partial{\cal M}_I$). One thus has $\widehat U_I=\widehat \Gamma_I+\widehat E_I$, where $\widehat \Gamma_I\in \widehat{\mathfrak{t}}$ is the connection of the principal ${\widehat{\mathfrak{t}}}$-bundle and $\widehat E_I\in \widehat{\mathfrak{hs}}(4)/\widehat{\mathfrak{t}}$ is a soldering one-form. The fields $(\widehat E_I,\widehat V_I,\widehat \Phi_I)$ are local representatives of sections of a ${\widehat{\mathfrak{t}}}$-bundle associated to the principle ${\widehat{\mathfrak{t}}}$-bundle.
The on-shell configurations $\left\{\widehat \Gamma_I,\widehat E_I,\widehat V_I,\widehat \Phi_I\,;\, \widehat T_I^{I'}\right\}_{\widehat{\mathfrak {t}}}$ form a moduli space $\mathfrak M_{\widehat{\mathfrak{t}}}$, which we refer to as the homotopy $\widehat{\mathfrak{t}}$-phase of the theory. 
The aim is to coordinatize this space using globally defined observables ${\cal O}_{\widehat{\mathfrak {t}}}[\widehat \Gamma,\widehat E,\widehat V,\widehat \Phi;\widehat T]$ built from the locally defined data. 
These are functionals that are \cite{Boulanger:2011dd,Sezgin:2011hq} i) manifestly $\widehat{\mathfrak{t}}$-invariant off shell; and ii) invariant under all canonical transformations of ${\cal C}$ on shell (hence in particular the diffeomorphisms of ${\cal M}$). 
Such quantities can be constructed by integrating manifestly $\widehat{\mathfrak{t}}$-invariant quantities built from $\star$-products of $(\widehat\Gamma,\widehat E,\widehat \Phi,\widehat V;\widehat T)$ over various submanifolds of the correspondence space ${\cal C}$; for example, in constructing $p$-form charges, one integrates on-shell de Rham closed densities over over nontrivial cycles $\Sigma\times {\cal Y}\times {\cal Z}$ where $\Sigma\subset {\cal M}$. 

There are two basic types of phases of the theory: the unbroken phase for which $\widehat {\mathfrak{t}}=\widehat{\mathfrak{hs}}(4)$ and hence $\widehat E=0$, and various broken soldered phases for which $\widehat {\mathfrak{t}}\equiv \widehat {\mathfrak{m}}\subset \widehat{\mathfrak{hs}}(4)$ and hence $\widehat E$ is nontrivial. In the latter case, one may refer to $\widehat {\mathfrak{m}}$ as a generalized Lorentz algebra, and on general grounds it is assumed that $\widehat {\mathfrak{m}}\supseteq {\mathfrak sl}(2,\Comp)$. There is no unique choice of $\widehat {\mathfrak{m}}$, however, so there exist many soldered phases; for further details, see \cite{Sezgin:2011hq}.

Let us mention briefly the basic features of the soldered phases and the topological phase.

\subsubsection*{Soldered phases}

A soldered phase $\mathfrak M_{\widehat{\mathfrak m}}$ is characterized by observables ${\cal O}_{\widehat{\mathfrak m}}[\widehat E,\widehat \Phi]$ that are manifestly $\widehat{\mathfrak m}$-invariant off shell and diffeomorphism invariant on shell (which is to say that they are intrinsically defined on associated $\widehat{\mathfrak m}$ bundles) where $\widehat E$ is a projection of $\widehat U$, referred to as the soldering one-form, that transforms homogeneously under gauge transformations valued in $\widehat{\mathfrak{m}}$.
Letting $\Sigma$ denote a nontrival closed cycle in the base manifold, one has (for further details, see \cite{Sezgin:2011hq}):
\begin{itemize}
\item Homotopy charges ${\cal Q}^R[\Sigma|\widehat E,\widehat \Phi]=\oint_{\Sigma}(\omega^R[\widehat E,\widehat \Phi]+K^R)$ where $\omega^R$ is a set of globally defined differential forms that are equivariantly closed on shell, \emph{viz.} $d\o^R+f^R(\o^S)\approx 0$ (using the equations of motion for the locally defined fields), and $K^R$ are the globally defined solutions to $dK^R=f^R(\o^S)$ on $\Sigma$;
\item Minimal areas ${\cal A}_{\rm min}[\Sigma|G_{(s)}]$ (and other brane observables appearing in non-topological brane partition functions) derived via norms induced by generalized metrics appearing as singlets in the symmetric direct products of the frame field; using commuting coordinates $X^M$ of dimension length on a Lagrangian submanifold and  letting $(\widehat \Phi;-,\dots,-)$ denote $s$-linear and totally symmetric $\widehat{\mathfrak m}$-invariant functions on the coset, one has the rank-$s$ metrics
\be (dX)^s|_{G_{(s)}}~=~dX^{M_1}\cdots dX^{M_s} G_{M_1\dots M_s}(X)~=~ \lambda^{-s} \eta_{(s)}(\widehat \Phi;\widehat E,\dots,\widehat E)\ ,\ee
where $\l$ is a massive parameter introduced such that $\widehat E$ and $G_{M_1\dots M_s}$ can be taken to be dimensionless.
\end{itemize}
Metric phases arise within soldered phases as the soldered form $\widehat E$ picks up vacuum expectation values. 
In the limit where the frame field vanishes as the Weyl zero-form is held fixed:
\be \widehat E~\rightarrow~0\ ,\quad \widehat \Phi~~{\rm fixed}\qquad \Rightarrow\qquad {\cal Q}[\widehat E,\widehat \Phi]\,,~~{\cal A}_{\rm min}[\Sigma|G_{(s)}]~\rightarrow~0\ ,\label{Elimit}\ee
the order parameters for the metric phase degenerate (vanish or diverge).
We stress once more the required status of the Weyl zero-form $\widehat \Phi$ as an independent field for the previous limit to make sense.

\subsubsection*{Unbroken topological phase}

The unbroken phase is characterized by observables that are manifestly $\widehat{\mathfrak hs}(4)$-invariant off shell and diffeomorphism invariant on shell (any such observable of course remains an observable in the various broken phases). The near-integrability of Vasiliev's higher-spin gravity motivates the following two types of observables in generally covariant systems which do not break any gauge symmetries (for further details, see \cite{Sezgin:2011hq}): 
\begin{itemize}
\item Locally accessible observables given by zero-forms ${\cal I}[p_0|\widehat \Phi]$ where $p_0$ is a point on the base manifold, obeying 
\be d{\cal I}[\widehat \Phi]~\approx~0\ ,\ee
which we refer to as zero-form invariants\footnote{\emph{c.f.} the zero-form invariants of higher-spin gravity introduced in \cite{Engquist:2005yt,Sezgin:2005pv} and  \cite{Sezgin:2005hf,Iazeolla:2007wt}; see also \cite{Boulanger:2008up} for a discussion of the r\^ole of massive parameters in constructing zero-form invariants and the notion of dual Weyl zero-form in strictly massless theories, and \cite{Giddings:2007nu} for a similar construction in the case of de Sitter gravity.}; \\
\item Multi-locally accessible observables ${\cal W}[\Gamma(p_1,\dots,p_n)|\widehat U,\widehat V,\widehat \Phi]$ depending on $\widehat \Phi$ and the one-form connections in $\widehat U$ and loops $\Gamma(p_1,\dots,p_n)\subset {\cal M}$ passing through $n$ special points,  such that ${\cal W}$ is independent under smooth deformations of the interiors of the loop and ($i=1,\dots,n$)
\be d_{p_i} {\cal W}~\approx~0\ ,\ee
which we refer to as decorated Wilson loops. 
\end{itemize}
In the absence of nontrivial monodromies in the flat connection on ${\cal M}$, the decorated Wilson loops collapse formally to the zero-form charges $\cI[p_0|\F]$ \cite{Sezgin:2011hq}. Their physical meaning will be discussed in Section \ref{Sec:3}, and the perturbative existence of one particular class of zero-form charges, based on the supertraces given in Eq. \eq{Inv1}, will be spelled out in Section \ref{Large-contour prescription}.

\subsection{Perturbative expansion in the Weyl zero-form}
\label{Expansion in the Weyl zero-form}

\subsubsection{Real-analytic master fields}

Using standard techniques, the twistor space equations \eq{middle} and \eq{e o m} can be solved locally in $Z$-space using a gauge function $\widehat \l$ for $\V$ and starting from an initial datum
\be U~=~\widehat U|_{Z=0}~\in~\mathfrak{hs}(4)\ ,\qquad \Phi~=~\widehat\Phi|_{Z=0}~\in~T[ \mathfrak{hs}(4)]\ ,\label{spacetimedata}\ee 
where the reduced adjoint and twisted-adjoint representations of the minimal-bosonic models are defined by
\be  \mathfrak{hs}(4)~=~\left\{~ \l(Y)~:~\tau(\l)~=~\l^\dagger~=~- \l~\right\}\ ,\qquad  \rho(\l) \l'~=~[\l,\l']_\star\ ,\ee 
\be T[ \mathfrak{hs}(4)]~=~\left\{~ C(Y)~:~\tau( C)~=~\pi(C)\ ,~ C^\dagger~=~\pi( C)~\right\}\ ,\qquad \rho(\l) C~=~[\l,C]_\pi\ .\ee 
The general perturbative form of the solution reads
\be \widehat\Phi~=~\widehat\Phi(\widehat\l,\Phi)\ ,\qquad \V~=~q\widehat\l+\V[{\widehat\l,\Phi}]\ .\ee
To fix the gauge function $\widehat\l$, one may impose the radial twistor gauge condition  
\be i_Z \widehat V~=~0\quad \Rightarrow\quad {\cal L}_Z\widehat\l ~=~ - i_Z \V_{\widehat\l,\Phi}\ ,\label{gaugecondition}\ee
where $i_Z$ and ${\cal L}_Z$ denote the inner and Lie  derivatives, respectively, along the radial vector field
\be  Z~=~ Z^\una \partial_\una\ .\ee 
The gauge condition  \eq{gaugecondition} leads to a unique and real analytic solution under the assumption that the initial datum $\F$ and $W$ are real analytic functions on ${\cal Y}$ with a $\Phi$-expansion into perturbative building blocks with well-defined $\star$-product compositions. 

To implement the gauge condition, one may use the homotopy contraction operator
\be \rho_\C~=~ i_Z \oint_\C {dt\over 2\pi i t} \gamma(t) \, t^{{\cal L}_Z}\ ,\quad \gamma(t)~=~\log {t\over 1-t}\ ,\label{j integration}\ee 
where $\gamma(t)$ is taken to branch along $[0,1]$ and $\C$ is a closed contour encircling $[0,1]$ counter clockwise. If $\widehat j=\frac1{p!} dZ^{\una_1}\cdots dZ^{\una_p} \widehat j_{\underline{\a_1\dots \a_p}}(Z)$ is a $p$-form of degree $p\geqslant 1$ that is real-analytic\footnote{The statement that a symbol is real-analytic is ordering dependent; see Section \ref{Obs} for a discussion.} and $q$-closed, then $ t^{{\cal L}_Z}\widehat j=\frac1{p!} dZ^{\una_1}\cdots dZ^{\una_p} ~t^{p}~\widehat j_{\underline{\a_1\dots \a_p}}(tZ)$ is a real-analytic function in $t$\,. It follows that $q(\rho_\C \widehat j)$ can be re-written by integrating by parts in $t$, which leaves no boundary term since $\C$ is closed, after which $\C$ can be deformed to a simple pole at $t=1$. One thus has (the last property is nontrivial only if $p=1$)
\be q \,(\rho_\C \,\widehat j)~=~\widehat j\ ,\qquad i_Z\,(\rho_\C \,\widehat j)~=~0\ ,\qquad (\rho_\C \,\widehat j)|_{Z=0}~=~0\ ,\label{toshow}\ee
for any closed homotopy contour $\C$ encircling $[0,1]$ counter clockwise. When acting on sufficiently regular twistor-space forms, the contour $\Gamma$ can be collapsed onto the branch-cut using the fact that \be \oint_{[0,1]} \frac{dz}{2\pi i} \gamma(t) f(t) ~=~\int_0^1 dt f(t)\ ,\label{collapse}\ee
for functions $f(t)$ that do not diverge faster than than $(t-x)^{y}$ for all $x\in [0,1]$ and some $y>-1$. Acting in the class of such functions, one has  \be\rho_\Gamma~=~\rho~:=~i_Z\frac1{{\cal L}_Z}\ ,\label{collapserho}\ee
and the homotopy contracting property follows immediately from  
\be q~\rho =1-i_Z\frac1{{\cal L}_Z}q\ .\ee

Returning to the twistor-space equations \eq{e o m}, they can thus be rewritten as 
\be  \F~=~\Phi-\rho_\C\left(\left[\V,\F\right]_\pi\right)\ ,\qquad \widehat V~=~-\rho_\C\left(\widehat\Phi\star\widehat J+\V\star \V\right)\ ,\ee 
under the assumption of the twistor gauge \eq{gaugecondition} and real-analyticity of the initial datum\footnote{The real-analycity properties of $\Phi(Y)$ leak over into $Z$-space via the application of the Klein operators.} $\Phi(Y)$\,.
These algebraic equations can then be solved iteratively in a perturbative expansion of the form  
\bea &\widehat \Phi~=~\sum_{n=1}^\infty\widehat\Phi^{(n)}[\Phi,\dots,\Phi]\ ,\qquad  
\widehat V~=~\sum_{n=1}^\infty \widehat V^{(n)}[\Phi,\dots,\Phi]\ ,&\label{iterative expansion}\eea
where $\F^{(n)}$ and $\V^{(n)}$ are $n$-linear symmetric functionals of $\Phi$
and \be \widehat \Phi^{(n)}|_{Z=0}~=~ \delta_{n1}\Phi\ .\ee 
One then applies the homotopy operator to \eq{middle}. Using $\rho_\C \V=0$\,, which implies $\rho_\C d\V=d(\rho_\C\V)=0$\,, one has 
\be  \U~=~ U-\rho_\C\left(\left[\V,\U\right]_\star\right)\ ,\ee 
with the perturbative solution 
\be \U~=~U+\sum_{n=1}\widehat U^{(n)}_1[U;\Phi,\dots,\Phi]~=~\left(1+\sum_{n=1}^\infty \widehat L^{(n)}\right)^{-1}U\ ,\label{U}\ee 
for homotopy operators $\widehat L^{(n)}\widehat f=\rho_\C\left(\left[\V^{(n)},\widehat f\right]_\star\right)$\,.

These solutions are formal in the sense that\footnote{Working more carefully one can also make active use working in intermediate alternative ordering schemes; see Section \ref{Obs}.} at the $n$-th level of the $\Phi$-expansion, Eqs. \eq{middle} and \eqref{e o m} imply
\be q\widehat\Phi^{(n)} ~\approx~ -\sum_{n_1+n_2=n} [\V^{(n_1)},\F^{(n_2)}]_\pi\ ,\qquad q\V^{(n)}~\approx~-\F^{(n)}\star \widehat J-\sum_{n_1+n_2=n} \V^{(n_1)}\star V^{(n_2)}\ ,\label{qequation}\ee
\be q\U^{n}~\approx~-d\V^{(n)}-\sum_{n_1+n_2=n} [\V^{(n_1)},\U^{(n_2)}]_\star\ ,\label{qequation2}\ee 
where by the perturbative assumption, the lower-order building blocks $\{\F^{(n')},\V^{(n')}\}_{n'=1}^{n-1}$ belong to an associative $\star$-product algebra and obey their respective equations of motion and  gauge condition. This implies that the right-hand sides in \eq{qequation} and \eq{qequation2} are $q$-closed. Thus, if the right-hand sides are in addition real-analytic \emph{after} the $\star$-products have been performed, then $\F^{(n)}$ and $\V^{(n)}$ can be obtained by applying $\rho_\C$ for any closed contour $\C$ enclosing $[0,1]$\,. This fact can be used to set up perturbative regularization methods, as we shall discuss in Section \ref{Master fields: perturbative associativity}.

\subsubsection{Residual $\mathfrak{hs}(4)$ gauge transformations}\label{sec gauge invariance}
\label{Residual gauge}

The physical gauge condition \eq{gaugecondition} is preserved by full gauge transformations \eq{gaugetransf} with residual gauge parameters obeying
\be i_Z\left(q\widehat\e+\left[\V,\widehat\e\,\right]_\star\right)~=~0\ ,\ee 
which can be rewritten using $i_Z\V=0$ and \eq{contractionZ} as
\be {\cal L}_Z\widehat \e+i \left[\V^\una,\partial^{(Y)}_{\una}\widehat\e\,\right]_\star -i \left\{\V^\una,\partial^{(Z)}_{\una}\widehat\e\right\}_\star~=~0\ .\ee 
Under the assumption of real-analycity in twistor-space, this implies the perturbative expansion
\be \widehat\e[\e;\Phi]~=~ \e+\sum_{n=1}^\infty\widehat\e^{(n)}[\e;\Phi,\dots,\Phi]\ ,\ee 
where $\widehat\e^{(n)}$\,, which are linear functionals of the $\mhs(4)$-valued gauge parameter $\e(Y)$\,, obey
\be {\cal L}_Z\widehat\e^{(n)}+i \sum_{n_1+n_2=n} \left[\V^{(n_1)\una},\partial^{(Y)}_{\una}\widehat\e^{(n_2)}\right]_\star -i \sum_{n_1+n_2=n} \left\{\V^{(n_1)\una},\partial^{(Z)}_{\una}\widehat\e^{(n_2)}\right\}_\star~=~0\ .\ee 
The induced residual $\mhs(4)$-transformations acting on the twisted-adjoint initial data are given by
\be  \delta_\e \Phi~=~(\delta_{\widehat\e}\widehat\Phi)|_{Z=0}~=~ -\left.\left[\widehat\e[\e;\Phi],\widehat\Phi[\Phi]\right]_\pi\right|_{Z=0}\ ,\label{res}\ee 
with softly deformed closure relations 
\be [\delta_{\widehat\e_1}, \delta_{\widehat\e_2}]  ~=~  \delta_{\widehat\e_{12}}\ ,\qquad \widehat\e_{12}~=~[[ \widehat\e_1,{\widehat\e_2}]+\delta_{\e_2}\widehat\e_1-\delta_{\e_1}\widehat\e_2\ ,\ee 
where $\delta_{\e_{1,2}}\widehat\e_{2,1}$ is the $\delta_{\e_{1,2}}\Phi$-variation of $\widehat\e[\e_{2,1};\Phi]$\,. Perturbatively,
\be \delta_\e \Phi~=~\sum_{n=0}^\infty \delta^{(n)}_\e \Phi\ ,\quad \delta^{(n)}_\e \Phi~=\delta^{(n)}[\e;\Phi,\dots,\Phi]\Phi~=~-\sum_{n_1+n_2=n}\left.\left[\e^{(n_1)},\widehat\Phi^{(n_2)}\right]_\pi\right|_{Z=0}\ ,\ee 
where the leading order is given by \be \delta_{\e}^{(0)}\Phi~=~-[\e,\Phi]_\pi\ .\ee

\subsection{Gauge function methods}

\subsubsection{General ideas and r\^ole of zero-form charges}

Given a graded exterior differential algebras, its locally defined solution space, including unbroken as well as broken gauge parameters (\'a la Cartan), can be sliced into orbits generated by gauge functions from reference solutions. In free cases, whether graded commutative or associative, the spaces of reference solutions can be taken to consist of constant zero-forms. In quasi-free cases, with additional algebraic constraints, these spaces acquire more structure. In particular, in graded associative systems with nontrivial central and closed terms in positive degrees, such as Vasiliev's equations, nontrivial reference solutions must contain fields with strictly positive form degree. 

In the Vasiliev system, a nonvanishing twisted-adjoint integration constant $C$ indeed implies a nontrivial reference solution $(\widehat \Phi'_C,\widehat V'_C)$ related to Wigner's deformed oscillator algebra and there are also nontrivial flat connections $\widehat V'_\theta$ in twistor space\footnote{Nontrivial flat connections can also arise on $T^\ast{\cal M}$.} for vanishing $C$ labeled by moduli parameters $\theta$; for examples in the case of the four-dimensional bosonic models, see \cite{Sezgin:2005pv,Sezgin:2005hf,Iazeolla:2007wt,WipCarlo}. The locally defined solution space to the full system \eq{unifold}--\eq{e o m} can thus be coordinatized reference solutions $(\widehat \Phi'_{C;\theta},\widehat V'_{C;\theta})$ and a master gauge function $\widehat L$ on the correspondence space ${\cal C}$. The solution $(\widehat \Phi'_C,\widehat V'_C;\widehat L$ can then be reduced down to submanifolds of ${\cal C}$, such as ${\cal Z}\times{\cal Y}$ and ${\cal M}\times{\cal Y}$, leading to dualities between formulations in the twistor space and spacetime, whose comparison furnishes one of the key motivations for seeking geometric formulations of higher-spin gravity.

Alternatively, as far as formulations on ${\cal M}$ are concerned, the full system can first be reduced down to ${\cal M}$, which yields free and graded commutative albeit perturbatively defined differential algebras, with locally defined solution spaces coordinatized in terms of $C$ and reduced gauge functions $L$. In principle, the resulting two approaches to formulations on ${\cal M}$ form a commutative diagram. Technically speaking, the path consisting of reducing $(\widehat \Phi'_C,\widehat V'_C;\widehat L$ is easier to implement than the one consisting of reducing the equation system. These two paths can be matched, however, at the level of classical observables, which is a second key rational for seeking geometric formulations of higher-spin gravity. 

In both of the above considerations, zero-form charges play a natural r\^ole: To begin with, these are natural basic observables of the aforementioned twistor-space formulation. Moreover, although generally requiring global considerations, as discussed in Section \ref{Global formulation: phases and observables}, the aforementioned two paths to formulations on ${\cal M}$ can actually be compared directly in a single coordinate chart using the zero-form charges, as we shall examine in more detail in Sections \ref{Sec:3} and \ref{Large-contour prescription}.

\subsubsection{Master gauge functions in correspondence space}

\label{Twistor-space picture}

Locally, the correspondence space ${\cal C}\stackrel{\rm loc}{\cong} T^\ast {\cal M}_I \times {\cal Z}\times {\cal Y}$, where ${\cal M}_I$ denotes a chart of ${\cal M}$. Eqs. \eq{unifold} and \eq{middle} can be integrated explicitly in $T^*{\cal M}_I$ using a gauge function $\widehat L_I$ \cite{Vasiliev:1990bu}; for further discussions of the gauge function method in the context of globally defined solutions, see \cite{Boulanger:2011dd,Sezgin:2011hq}, and for various applications, see \cite{Sezgin:2005pv,Iazeolla:2007wt,Giombi:2010vg}. Thus, suppressing the chart index, one has
\be \U^M~=~\widehat L^{-1}\star \partial^M \widehat L\ ,\qquad\widehat U_M~=~\widehat L^{-1}\star \partial_M \widehat L\ ,\label{gf1}\ee\be  \V_\una~=~\widehat L^{-1}\star (\partial_\una+\V'_\una)\star \widehat L\ ,\qquad \F~=~\widehat L^{-1}\star \F'\star\pi( \widehat L)\ ,\label{gf2}\ee 
where  the gauge function and the transformed master fields obey
\be \partial^M(\V'_\una ,\F')~=~0\ ,\qquad  \partial_M(\V'_{\una},\F')~=~0\ ,\ee 
\be 
 q\F'+\V'\star \F'-\F'\star\pi(\V')~=~0\ ,\quad q\V'+\V'\star\V' + \F'\star \J~=~0\ .\label{prime}
\ee
These equations are to be solved subject to initial conditions 
\be \F'|_{Z=0}~=~C\ ,\qquad  \widehat L|_{X=P=0}~=~1\ ,\label{twistordata}\ee
and boundary conditions on $\partial (T^\ast{\cal M})$ and in twistor space, \emph{i.e.} a choice of residual gauge function, say $L(X,P,Y)$, and flat twistor-space connections \cite{Sezgin:2005pv,Iazeolla:2007wt}, respectively. We assume that the latter are compatible with the twistor-gauge
\be i_Z \V~=~0\ .\label{twgauge}\ee
It follows that 
\be  \F'~=~ \F|_{X=0}\ ,\qquad \V'~=~ \V|_{X=P=0}\ ,\qquad i_Z \V'~=~0\ ,\ee 
such that 
\be C~=~\F'|_{Z=0}~=~\widehat \Phi_{X=P=Z=0}~=~ \Phi|_{X=P=0}\ ,\label{defC}\ee
where $\Phi$ is the reduced Weyl zero-form defined in \eq{spacetimedata}. The twistor-gauge condition implies that the gauge function obeys
\be \widehat L^{-1}\star {\cal L}_Z\widehat L+i(\partial^{\una}_{(Y)}+\partial^{\una}_{(Z)})\widehat L^{-1}\star (\partial^{(Z)}_{\una}+\V'_{\una})\star \widehat L+i \widehat L^{-1}\star  (\partial^{(Z)}_{\una}+\V'_{\una})\star (\partial^{\una}_{(Y)}-\partial^{\una}_{(Z)})\widehat L~=~0\ .\label{Leq}\ee
Under the assumption of real-analyticity, it follows that $\widehat L_{C;\theta;\l}=L\star \exp_\star(\widehat \l_{C;\theta;\l})$ where $\widehat \l_{C;\theta;\l}$ is a the particular solution to \eq{Leq}, that depends on $C$ and the moduli $\theta$ for the flat connection  in twistor space (see \cite{Sezgin:2005pv,Iazeolla:2007wt}), and $L=\exp_\star (\l)$, with $\l\in \mathfrak{hs}(4)$, is the homogeneous solutions representing the residual gauge degrees of freedom. 

The initial/boundary value problem is thus set up as follows: one first selects a twistor-space background $\V'_{0;\theta}$ and a residual gauge function $L=\exp_\star(\l)$ where $\l\in\mathfrak{hs}(4)/\mathfrak{m}$ and calculates the vacuum gauge function $\widehat L_{0;\theta;\lambda}$, describing a rigid higher-spin extension of $AdS(4)$. Into this vacuum configuration, local degrees of freedom are injected via the initial data $C$\,. The master fields $\F'_{C;\theta}$ and $\V'_{C;\theta}$ and the deformed gauge function $\widehat L_{C;\theta;L}$ are then obtained using either perturbative or exact methods, from which one can obtain the full albeit locally defined description in correspondence space via \eq{gf1} and \eq{gf2}. Reductions to ${\cal Z}$ and ${\cal M}_n$ then yield two dual descriptions which one may explore systematically using classical observables. 

In particular, taking trivial $\theta$ and reducing $T^\ast{\cal M}$ down to ${\cal M}_4$ and can choose $\l(x,Y)\in \mathfrak{so}(2,3)/\mso(1,3)$, describing an $AdS(4)$ vacuum, and expand $C(Y)$ in terms of twistor-space functions dual to localizable Weyl tensors $AdS(4)$. In the perturbative approach, Eq. \eq{prime} and is then solved using the homotopy contractor $\rho_\Gamma$ given in \eq{j integration}, that one may think of as propagators in twistor space. After having solved also \eq{Leq}, the  fields $\F'_{C;0}$ and $\V'_{C;0}$ can be mapped back to spacetime using the gauge function $\widehat L_{C;0;\l}$. This yields locally defined master fields $(U,\Phi)$ full of non-localities, and one may ask the question of whether locality is recuperated at the level of a suitable set of observables. For example, this has been found to be the case at the level of holographic three-point correlation functions \cite{Giombi:2009wh,Giombi:2010vg}. In this sense, the relatively intractable problem of dealing with the double perturbative expansion in spacetime is mapped to the more amenable problem of constructing associative algebra elements $(\widehat \l,\V',\F')$ in twistor space. 
In other words, the regularization of strongly coupled derivative expansions in spacetime is mapped to the arguably more tractable problem of regularizing $\star$-products in twistor space.

\subsubsection{Reduced gauge functions in spacetime}

Inserting the perturbative $\Phi$-expansions for $\widehat \Phi$ and $\widehat U$ given in \eq{iterative expansion} and \eq{U}, respectively, into \eq{unifold}, and restricting to $P_M=Z^{\underline\a}=0$, yields an unfolded description in terms of a free differential algebra on ${\cal M}$ of the form
\be d\Phi+P(U;\Phi)~\approx~0\ ,\qquad dU+J(U,U;\Phi)~\approx~0\ ,\label{PhiW}\ee
where $P$ and $J$ are linear and bilinear in $U$, respectively, and depend nonlinearly on $\Phi$. The quantity $Q:=P\cdot \frac{\partial}{\partial\Phi}+ J\cdot\frac{\partial}{\partial U}$ is a flow vector of degree one, acting in a graded targetspace with coordinates $(\Phi,U)$, and obeying the Cartan integrability condition $\{Q,Q\}=0$ (without further algebraic constraints on $(\Phi,U)$ so that \eq{PhiW} defines a free graded differential algebra). The locally defined solution spaces can be coordinatized using twisted-adjoint integration constant $C(Y)\in T[\mathfrak{hs}(4)]$ and adjoint gauge functions $\lambda(X,Y)\in\mathfrak{hs}(4)$, and expressed explicitly albeit perturbatively as\footnote{In general, the exponentiation may run into problems in the case of non-formal initial data, \emph{i.e.} initial data belonging to unitarizable representations whose elements in general are non-polynomial elements in the underlying associative algebra.}
\be (\Phi_{\lambda;C},U_{\lambda;C})~=~ \left[\exp(T_\lambda) (\Phi,U)\right]|_{U=0,\Phi=C}\ ,\label{CI}\ee
where $T_\lambda$ is a vector field of degree one in target space given by the generator Cartan gauge transformations, \emph{viz.} \be T_\lambda = -P(\lambda;\Phi)\cdot \frac{\partial}{\partial\Phi}+(d\lambda-2J(\lambda,U;\Phi))\cdot\frac{\partial}{\partial U}\ .\ee
As discussed in Section \ref{Global formulation: phases and observables}, the classical moduli spaces ${\cal M}_{\widehat{\mathfrak t}}$ consist of globally defined configurations of locally defined full master fields characterized by intrinsically defined classical observables ${\cal O}_{{\widehat{\mathfrak t}}}$ on $\widehat{\mathfrak{t}}$-bundles. Inserting the $\Phi$-expansions into these constructs, which may require integrating out $Z^\a$ and $P_M$, yields perturbatively defined $\mathfrak{t}$-invariant observables ${\cal O}_{\mathfrak t}(U,\Phi;T)$, where $\mathfrak t$ denotes the subalgebra of residual $\widehat{\mathfrak{t}}$-transformations, and $T$ denotes the reductions of the transition functions $\widehat T$ on $U$ and $\Phi$. Assuming that $\mathfrak t=\mathfrak m\subset \mathfrak{hs}(4)$, a nontrivial generalized Lorentz subalgebra, and letting $E\in \mathfrak{hs}(4)/\mathfrak{m}$ be the reduced soldering one-form, the order parameters for the soldered phase\footnote{We note that the physical data in $\l$ also contain monodromies, that are measured by Wilson loops, and described locally by integration constants of gauge functions associated with crossings between charts in the interior of ${\cal M}$.}
are manifestly $\mathfrak{m}$-invariant observables ${\cal O}_{\mathfrak m}(E,\Phi)$ such that ${\cal O}_{\mathfrak m}(E_{\l;C},\Phi_{\l;C})$ depend nontrivially on the boundary values $[\l]|_{\partial{\cal M}}$ of the generalized normal coordinates $[\l]\in \mathfrak{hs}(4)/\mathfrak{m}$. In the metric phase, the integration constant $C$ thus contain information of local deformations (including boundary states) while $[\l]|_{\partial{\cal M}}$ modulo ${\rm Diff}(\partial{\cal M})$ contain topological information of the background frame fields on $\partial{\cal M}$.

Two examples of order parameters are homotopy charges ${\cal Q}[\Sigma|E;\Phi]$ and minimal areas ${\cal A}[\Sigma|E;\Phi]$ \cite{Sezgin:2011hq}. These observables are given by integrals of manifestly $\mathfrak{m}$-invariant densities built from $E$ and $\Phi$ over $\Sigma\times{\cal Z}\times {\cal Y}$ where $\Sigma$ are topologically nontrivial cycles in ${\cal M}$. The densities are constructed such that the integrals are intrinsically defined, that is, depending only on the homology class $[\Sigma]$. One may refer to the order parameters as holographic observables in the sense that they remain invariant under diffeomorphims and homotopically trivial redefinitions of $\l$ which means that they localize to boundaries of ${\cal M}$ (or lower-dimensional submanifolds where other observables have already been inserted which one may think of as impurities).

In other words, from \eq{CI} it follows that starting from the initial data 
\be \Phi|_{p_0}~=~C\ ,\qquad \l|_{p_0}~=~0\ ,\label{p0}\ee
at the point $p_0\in{\cal M}$ where $T_\l$ vanishes, the unfolded field content can be constructed in a covariant Taylor expansion in the normal-coordinate directions. The local degrees of freedom in $C$ are thus measured by two dual sets of observables: they can be contracted into locally accessible zero-form charges ${\cal I}[p_0|\Phi]={\cal I}[p_0|C]$, or propagated boundaries where they become  boundary degrees of freedom measured by the holographic observables. In particular, as one can always set the gauge functions to zero inside a coordinate chart, it is always possible to gauge away the master gauge field $W$ in simply connected regions of ${\cal M}$ with compact support. 
In this sense, Eq. \eq{CI} manifests the fact that all local degrees of freedom arise via the Weyl zero-form \cite{Vasiliev:1988xc,Vasiliev:1988sa}, independently of whether the theory is free or interacting, and of the locality properties of various effective descriptions in metric phases 
\footnote{In \cite{Engquist:2002gy}, this local homotopy invariance of unfolded dynamics was used to derive the full superspace formulation of four-dimensional higher-spin supergravities.}. 

\subsubsection{Remarks on space-time reconstruction}

Finally, we wish to add a few more remarks on the interplay between geometry and algebra in unfolded dynamics.

In backgrounds with invertible vielbeins, the $Q$-cohomology, also known as $\sigma^-$-cohomology \cite{Vasiliev:2005zu}, can be contracted, leading to effective equations of motion in metric formulations. In particular, on ${\cal M}_4$, their free limits describe unitarizable local degrees of freedom. In the unfolded approach in general, there is, however, no \emph{a priori} reason why locality properties of the free equations of motion must persist at the level of interactions. In the interacting case, the physically relevant issue is rather whether the local degrees of freedom contained in the Weyl zero-form $\Phi$ exhibit localizability at the level of holographic or locally accessible observables. This issue may be studied either perturbatively in $\Phi$-expansion, as we shall do below in the case of locally accessible observables, or non-perturbatively provided that one has access to a fully non-linear sector of the moduli space; for examples, see \cite{WipCarlo}.

Another more algebraic way of reasoning is as follows: The full master fields $\widehat \Phi$ and $\widehat A$ are locally defined differential forms on the base manifold ${\cal B}$ taking their values in spaces of functions on ${\cal Y}$. 
One may think of such spaces as modules for a reduced higher-spin Lie algebra $\mhs(4)\supset \mso(2,3)$ as follows: the algebra $\mhs(4)$ itself is a subspace of the $\star$-commutator closed space of arbitrary polynomials on ${\cal Y}$. Various twisted-adjoint representations, containing the initial data $C$ defined by \eq{initialdata} (or \eq{initialdata2}), can then be generated starting from reference elements given by functions on ${\cal Y}$ that may in general be non-polynomial; among these one finds, for example, various unitarizable representations \cite{Iazeolla:2008ix}.   
This raises the issue on what physical grounds the theory selects its perturbative spectrum, that is, the set of admissible twisted-adjoint representations.
The systematic way of proceeding is thus to exploit the constraints on $C$ that arise from demanding well-defined classical observables, and in particular, well-defined zero-form charges.

\section{Zero-form charges and twistor-space quasi-amplitudes}\label{Sec:3}

In this Section we discuss formal properties of classical observables given by integrals over the doubled twistor space ${\cal Y}\times {\cal Z}$ evaluated at a single point in $T^\ast{\cal M}$. These observables are thus given by on-shell closed zero-forms on $T^\ast{\cal M}$ with a dual interpretation as basic blocks for amplitudes in twistor space; for further details on the latter interpretation, see the forthcoming paper \cite{Sezgin:2011hq}. In the next Section we shall then look in more detail into the regularization of these twistor-space quasi-amplitudes.

\subsection{Locally accessible observables and localizability of states}

An observable can be said to be locally accessible if it is nontrivial on shell in a single coordinate chart of spacetime.  
In unfolded dynamics, such an observable is a composite zero-form ${\cal I}(\Phi)$ that is closed on shell, \emph{i.e.} $d{\cal I}\approx0$\,. This is equivalent to that ${\cal I}$ is invariant under general Cartan gauge transformations\footnote{Consider an unfolded system with zero-forms $\Phi^i$ obeying $d\Phi^i+Q^i(\Phi;U)=0$ with $Q^i=U^r Q^i_r(\Phi) $ where $U^r$ denote the one-forms of the system. Under one-form gauge transformations with parameters $\e^r$\,, the zero-forms transform as $\delta_\e \Phi^i=-\e^r Q^i_r$\,. If ${\cal I}[\Phi]$ is closed on shell, that is $0=d{\cal I}=-U^r Q^i_r \partial_i {\cal I}$ for all $U^r$\,, then it follows that $\delta_\e {\cal I} =-\e^r Q^i_r \partial_i {\cal I}=0$ for all $\e^r$ as well. }, \emph{i.e.} $\delta_\e{\cal I}=0$. Hence ${\cal I}$ is a globally defined zero-form on the base manifold. 
From \eq{CI} and \eq{p0} it follows that  
\begin{equation}
 {\cal I}(\Phi_{\l,C})~=~{\cal I}(C) \ ,
\end{equation}
which hence defines a Casimir invariant for the unfolded system, or a generalized central charge. Alternatively, one may think of ${\cal I}$ as a building block for the quantum effective action on shell (including deformations and counter terms and evaluated with initial data prescribed by $C$)\footnote{For further discussions of on-shell actions in the context of unfolded dynamics, see \cite{Boulanger:2011dd,Sezgin:2011hq}. }.
In the curvature expansion, 
\begin{equation}
 {\cal I}(\Phi)~=~\sum_n {\cal I}^{(n)}(\Phi,\dots,\Phi)\ ,\qquad 
\delta_\epsilon \Phi~=~i\sum_{n=0} \rho^{(n)}(\epsilon;\Phi,\dots,\Phi)
\,\Phi\ ,\label{I(n)}
\end{equation}
where $\rho^{(0)}(\e)\Phi$ is a rigid gauge transformation, implying the 
equivariance relations  
\begin{equation}
 \sum_{n_1+n_2=n}n_1{\cal I}^{(n_1)}\left(\r^{(n_2)}(\epsilon;\Phi,\dots,\Phi)\Phi,
 \Phi,\dots,\Phi\right)=0 \ .
\end{equation}
In particular, one has zero-form charges ${\cal I}_\NumbCurv$ such that 
\begin{equation}
\cI^{(n)}_\NumbCurv(\Phi,\dots,\Phi)~=~0~\mbox{ for $n<\NumbCurv$}~~\Rightarrow~~
\cI^{(\NumbCurv)}_\NumbCurv\left(\delta^{(0)}_\e \Phi,\Phi,\dots,\Phi\right)~=~0\ .
\end{equation}
whose leading terms are thus invariants of the rigid nonabelian gauge algebra.

In general, the twisted-adjoint module $T$ of the unfolded system decomposes into representations $T\hspace{-5pt}\downarrow_{\mathfrak h}$ consisting of states labeled by quantum numbers $s$ of various subalgebras $\mathfrak{h}\subset \mathfrak{hs}(4)$, such as for example unitarizable one-particle states or solitons; for a discussion in the context of higher-spin gravity, see \cite{Iazeolla:2008ix}. In such a sector, one may choose a reference state $f^{s_0}_{p_0}\in T\hspace{-5pt}\downarrow_{\mathfrak h}$ associated to the base point $p_0$. This state can then be rotated into a multiplet $\{f^s_{p_0}\}$ by $\mathfrak{h}$\,. These states are then translated by a gauge function $L_{q,p_0}$, obeying $L_{p,q}L_{q,r}=L_{p,r}$, into
\begin{equation}
 f^s_q~=~\rho^{(0)}(L_{q,p_0}) f^s_{p_0}\ ,\label{q0}
\end{equation}
where $\rho^{(0)}$ denotes the representation matrix in $T\hspace{-5pt}\downarrow_{\mathfrak h}$. The initial data 
\be \Phi|_{p_0}~=~C~=~C\hspace{-5pt}\downarrow_{\mathfrak h}~=~\sum_{i} f^{s_i}_{q_i} C_i\ ,\ee
where $C_i\in \Comp$, describes a free solution 
\be \Phi^{(1)}(p)~=~\rho^{(0)}(L^{-1}_{p,p_0})C\hspace{-5pt}\downarrow_{\mathfrak h}~=~\sum_{i} \Phi^{s_i}_{q_i}(p) C_i\ ,\ee
where $\Phi^{s_i}_{q_i}(p)=\rho^{(0)}(L_{p_0,p}L_{q_i,p_0})f^{s_i}_{p_0}$ obeys $\Phi^{s_i}_{q_i}(q_i)=f^{s_i}_{p_0}$, the idea being that this is the dominant contribution to $\Phi^{(1)}(p)$ at $p=q_i$.  In general, the perturbative zero-form invariants can be decomposed as 
\begin{equation}
 {\cal I}^{(n)}(C\hspace{-5pt}\downarrow_{\mathfrak h})~=~\sum_{k} {\cal I}^{(n),k}(C\hspace{-5pt}\downarrow_{\mathfrak h})\ ,\ee\be 
 {\cal I}^{(n),k}(C\hspace{-5pt}\downarrow_{\mathfrak h})~=~\sum_{\tiny\ba{c}i_1,n_1;
\dots;i_k,n_k\\n_1+\dots+n_k=n\ea}\left(\prod_{l=1}^k (C_{i_l})^{n_l}\right)  
\widetilde {\cal I}^{~(n),k}_{i_1,n_1;\dots;i_k,n_k}\ ,
\end{equation}
where 
\begin{equation}
\widetilde{\cal I}^{~(n),k}_{i_1,n_1;\dots;i_k,n_k}~=~ 
{\cal I}^{(n)}\left((f^{s_{i_1}}_{p_{i_1}})^{n_1},\dots,(f^{s_k}_{p_{i_k}})^{n_k}\right)\ , 
\end{equation}
that vanishs trivially if $k>n$\,. 
One can then say that the sector in question exhibits locality and that 
$\Phi^s_q(p)$ is localized at $p=q$ if ${\cal I}^{(n)}(C\hspace{-5pt}\downarrow_{\mathfrak h})$ exhibit classical cluster decomposition in the sense that there is a hierarchy such 
that\footnote{This form of locality holds for spherically symmetric 
solutions of the four-dimensional Vasiliev system \cite{Didenko:2009td} (see \cite{WipCarlo} for an analysis of localizability) suggesting that also solitons 
play a r\^ole as ``perturbative'' building blocks in higher-spin gravity.} 
\begin{equation}
 \widetilde{\cal I}^{~(n),1}~\gg~
\widetilde{\cal I}^{~(n),2}~\gg~\cdots~\gg~\widetilde{\cal I}^{~(n),n}\ ,
\end{equation}
in limits where all positions are separated ``well enough'' in the background metric of
the gauge function; in other words, given a unitarizable sector, one may use the condition of cluster decomposition as a definition of what ``space-like'' separation should mean, though we shall not go into these details much further in this paper.

In sectors exhibiting locality, it is thus meaningful to think of $f^{s}_p$ as describing localized objects, independently of whether the effective equations of motions in the 
metric phase contain nonlocal interactions or not. One can then interpret the most 
separated (and smallest) pieces 
\be \widetilde {\cal I}^{~(n)}((p_1,s_1),\dots,(p_n,s_n))~:=~{\cal I}^{(n)}\left(f^{s_1}_{p_1},\dots,f^{s_n}_{p_n}\right)\ee 
as candidate building blocks for gauge-equivariant $n$-particle scattering amplitudes, which we refer to as quasi-amplitudes.

\subsection{Implementation in Vasiliev's higher-spin gravity}
\label{Traces and observables for unbroken and metric phases}

The $\star$-product algebra of functions on ${\cal Y}\times {\cal Z}$ admits a chiral trace operation given by
\be  \widehat{\rm Tr}[\widehat O(y,\yb;z,\zb) ]~=~\int_{\cal R} \frac{d^2y d^2\yb d^2z d^2\zb}{(2\pi)^4} ~\widehat f_O(y,\yb;z,\zb)\ ,\label{chiraltrace}\ee 
where $\widehat O(y,\yb;z,\zb)$ is an operator represented by the symbol $\widehat f_O$ in some ordering scheme, and integration domain ${\cal R}$ taken as in \eq{chiraldomain}. Formally, this trace operation is cyclic, independent of the choice of ordering scheme\footnote{The independence of ordering prescription is a consequence of the fact that a slight altering of prescription induces a change in the symbol given by a total derivative on ${\cal Y}\times {\cal Z}$ leading to boundary terms that vanish if the trace is finite. Strictly speaking, this argument requires symbolizable and universal orderings; for a more detailed discussion, see for example \cite{WipCarlo}.  } and obeying 
\be \widehat{\rm Tr}[\widehat f])^\dagger~=~
\widehat{\rm Tr}[\widehat f^\dagger]\ .\ee
It is thus natural to construct classical observables for $\mathfrak M_{\widehat{\mathfrak t}}$ using traces. Manifest $\widehat{\mathfrak t}$ gauge invariance off shell can be achieved by tracing $\star$-product composites built from the locally defined data $\{\widehat\Gamma_I,\widehat E_\charti,\V_I,\F_\charti\,;\,\widehat T_\charti^{\charti'}\}$ that transform in the adjoint representation of $\widehat{\mathfrak t}$. Invariance under canonical transformations of ${\cal C}$ on shell requires more detailed constructions.  

In the unbroken phase, a natural set of intrinsically defined observables are decorated Wilson loops in the Lagrangian submanifold ${\cal M}$ given by\footnote{For further details, see \cite{Sezgin:2011hq}.} 
\be {\cal W}\left[\Gamma(p_1,\dots,p_n)|\left\{\F_\charti,\U_\charti;\widehat T_\charti^{\charti'}\right\}\right]~=~ {\cal N}\,
\widehat{\rm Tr} \left[ {\rm P}_\Gamma \left\{ \prod_{\charti=1}^M \widehat\Psi_{m_\charti,\bar m_\charti ; n_\charti}|_{p_\charti}\exp_\star \left(\int_{\Gamma_{\charti}} \widehat 
U_{\charti}\right) \widehat T_{\charti}^{\charti+1} \right\}\right]\ ,\label{WL}\ee
where ${\cal N}$ is a normalization chosen such that the leading order in the curvature expansion is finite; the symbol P$_\Gamma$ denotes the path order along a path $\Gamma\subset {\cal M}$; and 
\be  \widehat\Psi_{m,\bar m;n}~=~\widehat\Psi^{\star m}\star \widehat{\overline\Psi}{}^{\star \bar m}\star ( \widehat\kappa\widehat{\bar\kappa})^{\star n}\ ,\ee
are adjoint impurities, where have defined \be \widehat \Psi~=~ \F\star \widehat\kappa\ ,\qquad \widehat{\overline\Psi}~=~(\widehat\Psi)^\dagger~=~\F\star\widehat{\bar\kappa}\ ,\label{defPsi}\ee 
and $m, \bar m, n\in \{0,1,2,\dots\}$ modulo the relations $( \widehat\kappa\widehat{\bar\kappa})^{\star 2}=1$ and  \be \widehat\Psi\star \widehat\kappa\widehat{\bar\kappa}~=~\widehat\kappa\widehat{\bar\kappa}\star \widehat\Psi~=~\widehat{\overline\Psi}\ ,\qquad \widehat \Psi^{\star 2}~=~\widehat {\overline \Psi}{}^{\star 2}\quad \Rightarrow\quad \widehat \Psi\star \widehat{\overline\Psi}~=~\widehat{\overline\Psi}\star \widehat{\Psi}\ .\ee
Viewed as a function of a fixed $p_i$ with the remaining impurities held fixed, this observable behaves as zero-form that is closed on shell.
Assuming that the decorations can be pushed together to a single point and that the loop is trivial (or if there are no monodromies), the Wilson loop can be expended in terms of zero-form charges ${\cal N}\widehat{Tr}\left[\widehat\Psi_{m,\bar m;n}\right]$.  Splitting into real and imaginary parts, one has ($\NumbCurv=2,4,\dots$)\footnote{The zero-form charges generate interaction ambiguities \cite{Sezgin:2011hq} and on-shell actions and related tree-amplitudes in twistor space \cite{Sezgin:2011hq} (see Appendix B). In both these applications, the zero-form charges appear together with free parameters that can be taken to contain the required normalizations. } 
\bea \label{Inv1}\cI_{\NumbCurv}&=&~{\cal N}_K\widehat{\rm Tr}[(\widehat\Phi\star\pi(\widehat\Phi))^{\star \NumbCurv}\star \widehat{\kappa} \widehat{\bar{\kappa}}]\ ,\label{IN}\\[5pt]\label{Inv2}\cI^\pm_{\NumbCurv+1}&=&{\cal N}^\pm_{K+1}~ \widehat{\rm Tr}[(\widehat\Phi\star\pi(\widehat\Phi))^{\star \NumbCurv}\star \F\star\frac12(\widehat \kappa\pm \widehat{\bar{\kappa}})]\ ,\\[5pt]\label{Inv3}
\cI'_{\NumbCurv}&=&{\cal N}'_K ~\widehat{\rm Tr}[(\widehat\Phi\star\pi(\widehat\Phi))^{\star \NumbCurv}]\ .\eea
These zero-form charges are on-shell closed and globally defined functionals. Two local representatives, say $(\widehat A_\charti,\F_\charti)$ and $(\A_{\charti'},\F_{\charti'})$, can be glued together with a transition function $\widehat T_{\charti}^{\charti'}$ only if ${\cal I}[\widehat\Phi_\charti]={\cal I}[\widehat\Phi_{\charti'}]$ for all zero-form charges ${\cal I}$. In the curvature expansion \eq{I(n)}, \emph{viz.} ${\cal I}[\F(C)]=\cI(C)=\sum_n {\cal I}^{(n)}(C,\dots,C)$ where the twisted-adjoint initial data $C$ is defined in \eq{twistordata}, one may ask whether the $n$-linear bose-symmetric functionals $ {\cal I}^{(n)}$ are well-defined. 
To normalize the leading order, one notes that the insertions of inner Kleinians localizes the chiral trace operation, as can be seen by going to overall Weyl order where 
\be \left[\widehat{\kappa}\right]_{\rm Weyl}~=~(2\pi)^2 \delta^2(y)\delta^2(z)\ ,\qquad \left[\widehat{\bar{\kappa}}\right]_{\rm Weyl}~=~(2\pi)^2 \delta^2(\yb)\delta^2(\zb)\ .\ee
One is thus led to the following normalizations
\be {\cal N}_K~=~1\ ,\qquad {\cal N}^\pm_{K+1}~=~{\cal N}\ ,\qquad {\cal N}'_K~=~{\cal N}^2\ ,\ee
where ${\cal N}$ is the inverse of the volume of chiral twistor space\footnote{One way of regularizing \eq{calN} is to interpret the integral over chiral twistor space as the trace of the identity operator in a Fock space. Alternatively, one may choose to work with subtractive regularization schemes. We leave both these interesting problems for future studies.}, 
\be {\cal N}^{-1}~=~\int \frac{d^2z}{2\pi}\ .\label{calN}\ee
With these choices, one has the leading terms
\bea \label{I1}\cI_{\NumbCurv}^{(\NumbCurv)}&=&{\rm STr}_y \,{\rm STr}_{\yb}\left[( C\star\pi(C))^{\star \NumbCurv}\right]\ ,\\[5pt]\label{I2}
 \cI_{\NumbCurv+1}^{ \pm (\NumbCurv+1)}&=&\frac12 ({\rm STr}_y {\rm Tr}_{\bar y}\pm {\rm Tr}_y \,{\rm STr}_{\yb}) \left[(C\star\pi(C))^{\star \NumbCurv}\star C\right]\ ,\\[5pt]\label{I3}\cI_{\NumbCurv}^{ \prime (\NumbCurv)}&=&{\rm Tr}_y {\rm Tr}_{\yb} \left[(C\star\pi(C))^{\star \NumbCurv}\right]\ ,\eea
where ${\rm STr}_y[f(y)]=f|_{y=0}$ and ${\rm Tr}_y[f(y)]=\int \frac{d^2y}{2\pi} f(y)$ \emph{idem} $\yb$.
As discussed in Section \ref{Twistor-space picture}, the harmonic expansion of free fields subject to boundary conditions in spacetime leads to initial data of the form\footnote{For further detailed information on the fiber approach to harmonic analysis, see \cite{Iazeolla:2008ix,Boulanger:2008up,Boulanger:2008kw}.  Essentially, the idea is that taking $C=f_{\Sigma;\L}$ yields a full zero-form $\F(X,P,Z;Y)$ with projections $\Phi(x,P=0;Y)=\F(x,P=0,Z=0;Y)$ and $\F'(Z;Y)=\F(X=0,P=0,Z;Y)$ obeying dual boundary conditions in spacetime and twistor space, respectively. }:
\be  C(Y)~=~\sum_{\Sigma} C_\Sigma(Y)\ ,\qquad C_\Sigma(Y)~=~ \sum_\L f_{\Sigma;\L}(Y) C_{\Sigma;\L}\ ,\label{C(Y)}\ee 
where $\Sigma$ labels different sectors of boundary conditions; $f_{\Sigma;\L}$ denote basis elements of these sectors, labeled here by some index $\Lambda$, and which one may think of as vertex operators; and the expansion coefficients $C_{\Sigma;\L}$ are commuting numbers. 
The corresponding $\mathfrak{hs}(4)$-invariant quasi-amplitudes are defined by
\be \widetilde{\cal I}(\{\Sigma_i,\L_i\}_{i=1}^{K})~=~ \frac 1{K!} \sum_{\tiny\ba{c}\mbox{permutations}\\[-5pt] \sigma\ea}{\rm STr}_{y}{\rm STr}_{\bar y}\left[f_{\Sigma_{\s(1)};\L_{\s(1)}}\star \cdots \star \pi(f_{\Sigma_{\s(K)};\L_{\s(K)}})\right]\ ,\ee
\be \widetilde{\cal I}^\pm(\{\Sigma_i,\L_i\}_{i=1}^{K+1})~=~\frac1{(K+1)!}\sum_{\tiny\ba{c}\mbox{permutations}\\[-5pt] \sigma\ea} \frac12({\rm Tr}_y {\rm STr}_{\bar y}\pm {\rm STr}_y {\rm Tr}_{\bar y})\left[f_{\Sigma_{\s(1)};\L_{\s(1)}}\star \cdots \star f_{\Sigma_{\s(K+1)};\L_{\s(K+1)}}\right]\ ,\ee
\be \widetilde{\cal I}^{\prime}(\{\Sigma_i,\L_i\}_{i=1}^{K})~=~\frac1{K!}\sum_{\tiny\ba{c}\mbox{permutations}\\[-5pt] \sigma\ea} {\rm Tr}_{y}{\rm Tr}_{\bar y}\left[f_{\Sigma_{\s(1)};\L_{\s(1)}}\star \cdots \star \pi(f_{\Sigma_{\s(K)};\L_{\s(K)}})\right]\ .\ee
Breaking $\mathfrak{hs}(4)$ down to $\mathfrak{sp}(4)$ one obtains quasi-amplitudes for fixed Lorentz spins. For example, one may label the basis elements by points $x \in AdS(4)$ and Lorentz spins $(s,s)$, \emph{viz.}
\be 
f_x^{(s.s)}~ = ~(L(x))^{-1}\star f^{(s,s)}_0\star \pi(L(x))\ ,\label{locstate}\ee
where the gauge function $L(x)$ is a coset coordinatization of $AdS(4)$ and $f^{(s,s)}_0$ can be taken to belong to various representations of the twisted-adjoint spin-$s$ module. This yields $K$-point twistor-space amplitudes labeled by $K$ bulk points; for example, the three-point functions
\be \widetilde \cI^\pm(1,2,3)~=~ \frac14(Tr_y STr_{\bar y}\pm STr_y Tr_{\bar y})\left[f^{(s_1,s_1)}_0\star \pi(L(x_1,x_2))\star \pi(f^{(s_2,s_2)}_0)\right.\nonumber\ee\be\hspace{5cm}\left. \star L(x_2,x_3)\star f^{(s_3,s_3)}_0\star \pi(L(x_3,x_1))\right]+(1\leftrightarrow 2)\ ,\ee
where the vertex operators are linked together by $L(x_i,x_{i+1})=L(x_i)\star L^{-1}(x_{i+1})$ ($x_4\equiv x_1$).
Using vertices that are localized at $x=0$ and sending $x_i$ to the boundary, one may examine whether the amplitudes fall off with evanescent terms consisting of $\mathfrak{hs}(4)$-invariant amplitudes, fixed entirely by kinematics. We leave this study, in particular the comparison with ordinary holographic amplitudes, for future studies.

\section{Regularization of twistor-space quasi-amplitudes}
\label{Large-contour prescription}

\subsection{Motivation and summary of results}

At the level of the perturbative expansion of the master fields, the $\star$-products on the right-hand sides of \eq{qequation} may produce singularities in the homotopy integration variables for particular initial data $C$. At the level of classical observables, additional potential divergencies may arise as the result of trace operations.
Thus, in order to calculate classical observables, one needs to set up regularization schemes, that in general may depend on the initial data $C$\,. Below, we first assemble a scheme based on analytical continuation of the closed contour $\Gamma$ in the homotopy operator $\rho_\Gamma$ in \eq{j integration}. We then apply this scheme to the calculation of twistor-space amplitudes in the sector of twistor-plane waves \cite{Vasiliev:1990bu,Bolotin:1999fa}. We recall that in this sector the master fields have well-defined perturbative expansions \cite{Vasiliev:1990vu} (see also \cite{Prokushkin:1998bq}), making it a convenient testing ground for the formalism. 

In the aforementioned context, we wish to remark on the following subtlety: In the sector of twistor plane waves, the closed-contour homotopy contractor $\rho_\Gamma$ collapses to the open-contour contractor $\rho$ as in \eq{collapserho}, leading to the presentation of the master fields in this sector given originally in \cite{Vasiliev:1990vu}. However, as one traces strings of master fields from this sector, singularities appear if one uses $\rho$, while no singularities appear if one uses $\rho_\Gamma$ with sufficiently large contour $\Gamma$. In the latter case, one can exchange the order of integration between the trace and the homotopy contractions, and evaluate the integrals over $\Gamma$ using residues. 

Proceeding in this fashion, our finding is that in the sector of twistor plane waves there exists at least one good set of zero-form charges and corresponding quasi-amplitudes, namely those in \eq{IN} based on supertraces on both chiral and anti-chiral sides. Interestingly enough, in this case, the residues in the first sub-leading order cancel due to twistor identities. One thus remains with the interesting possibility that all residues vanish, in which case these zero-form charges would be given by their leading contributions in \eq{I1}, though we have not found any general argument for such a protection mechanism. 

\subsection{Large-contour prescription and perturbative associativity}

\label{Master fields: perturbative associativity}
In order to give a prescription for the perturbative expansion of the master fields themselves one has essentially to make sure that it abides by the requirement of associativity.

Assuming that the initial data $C$ is such that there are no branch cuts ending at infinity\footnote{This holds for twistor-space plane waves.}, we propose to avoid any singularities at finite locations by taking all homotopy integration contours large. This prescription is compatible with integrability if there are no singularities appearing inbetween large homotopy contours as one performs the $\star$-products in the associators of the set $\{\F^{(n')},\V^{(n')}\}_{n'=1}^{n-1}$\,. 

To spell out the large-contour prescription in more detail, we define the ordered set \be \{\widehat f_i\}_{i=0}^\infty~=~\{\widehat J,\F^{(1)},\V^{(1)},\F^{(2)},\V^{(2)},\dots\}\ ,\ee 
the equations \eq{e o m} can be expanded as \be q\widehat f_i~=~C_{i}^{j,i-j} \widehat f_j\star \widehat f_{i-j}\ ,\label{qfi}\ee 
where $C_i^{j,i-j}=0$ if $j=0$ or $j=i$\,.  Using $\widehat f_0\star \widehat f_i=\pi(\widehat f_i)\star \widehat f_0$\,, one can show that the integrability amounts to that \be C_i^{j,i-j}C_{j}^{k,j-k}\left[\widehat f_{i-j};\widehat f_k;\widehat f_{j-k}\right]~=~0\ ,\label{cc}\ee where the associators \be \left[\widehat f_k;\widehat f_l;\widehat f_m\right]~= ~(\widehat f_k\star\widehat f_l)\star \widehat f_m-\widehat f_k\star(\widehat f_l\star \widehat f_m)\ .\ee 
Repeated homotopy integration yields $\widehat f_i$ $(i\geqslant 2)$ as $i-1$ closed-contour integrals, \emph{viz.}
\be \widehat f_i(\C_{i-1};\C_{i-2},\dots,\C_1)~=~\rho_{\C_{i-1}}\left(C_i^{j,i-j} \widehat f_j(\C_{j-1};\C_{j-2},\dots,\C_1)\star \widehat f_{i-j}(\C_{i-2};\C_{i-3},\dots,\C_j)\right)\ ,\label{ficontours}\ee
enclosing $[0,1]$ counter clockwise in accordance with the large-contour prescription:  in order to $\star$-multiply $\widehat f_i$ with other perturbative building-blocks from the left or the right --- for example in going higher up in perturbation theory or in looking at observables --- such $\star$-products shall be performed taking all contours in $\widehat f_i$  to be large, including the last contour $\C_{i-1}$\,. In particular, this prescription applies to $q\widehat f_i$\,, since
\be q\widehat f~=~ \widehat q\star \widehat f-(-1)^{\widehat f} \widehat f\star \widehat q\ ,\qquad \widehat q~=~\frac{i}2 dZ^{\una}Z_{\una}\ .\ee 
Thus, it follows that $q\widehat f_i$ is given by a large contour integral along $\C_{i-1}$ that can be deformed back to the poles coming from $\gamma(t_{i-1})$ as to reproduce \eq{qfi} if the $\star$-products on the right-hand side of \eq{ficontours} yield real-analytic symbols in $Z$-space when performed using the large-contour prescription. Applying $q$ once more one finds the integrability condition \eq{cc}, which one may as well --- having in mind composite operators --- replace by the stronger condition that $\{\widehat f_i\}$ form an associative algebra, \emph{viz.}
\be\left[\widehat f_k;\widehat f_l;\widehat f_m\right]~= ~0\ ,\label{associators}\ee
provided all $\star$-products are performed using the large-contour prescription. These associators vanish if there are no poles in between the nested large contours. 

\subsubsection*{Exchanging traces and large-contour homotopy integrals}\label{Obs}

Suppose that $\{\Phi^{(n)},\V^{(n)}\}_{n=1}^\infty$ is perturbative solution to the internal Vasiliev equations \eq{qequation} obtained using the large-contour prescription (that is, the generates an associative algebra and its members obey \eq{qequation}).  We then define the zero-form observables in \eq{Inv1}--\eq{Inv3} by first expanding perturbatively and then exchanging the traces with the closed homotopy integrals. On general grounds, we expect such multiple closed-contour integrals to be finite and given by residues at infinity, leaving the trivial possibility that these residues vanish identically for generic solutions, which motivates taking a closer look at a simple case\footnote{\label{tracefootnote}
We wish to stress that in symbol calculus, as the ordering changes, the nature of the symbol of a given operator, thought of as a function or a distribution, may change drastically; for example, real-analytic Gaussian symbols may turn into delta functions. Along such deformations, which form paths in the affine space of orderings, that is, the space of bases for the operator algebra, a typical phenomenon is that there appear ordering dependent singularities that start moving across the auxiliary $t$-planes. As these pass under the closed homotopy contours, the perturbatively defined master fields pick up finite residues that hence depend not only on the ordering but also the details of how the contours are drawn. These ambiguities are spurious, however, since they drop out from the traces used in defining observables, which are ordering-independent.}.

\subsection{The twistor plane-wave sector}
\label{plane waves}

\subsubsection{Definition}

We define the twistor plane-wave sector as the set of perturbatively defined solutions for which i) the expansion \eq{C(Y)} of the initial data $C$ is in terms of twistor-space plane waves, \emph{viz.}
\be\label{Lintegration}
C_i(Y)~=~ \int d^4\Lambda~ f_\L~ C_i(\L)\ ,\qquad f_\L~= ~ e^{i Y^\una \L_\una}~=~  e^{i (y \l + \yb \bar{\l} ) }\ ,
\ee
where $ \L_\una=(\l_\a,\lb_{\ad})$ are commuting twistor-space momenta; and ii) the $\Lambda$-integrals commute to the twistor-space integrals that arise in the perturbative expansion of master fields and classical observables. As we shall see, the second part of this definition requires the closed-contour version $\rho_\Gamma$ of the homotopy contractor defined in \eq{j integration}. 
Thus, in this sector we have
  \bea
 \label{full fourier expansion}
 \F^{(n)}[C,\dots,C]&=&  \int d^4\L_1\cdots d^4\L_n \widetilde\Phi^{(n)}_{\L_1,\dots, \L_n} C_{\L_1}\cdots C_{\L_n}\ ,\\[5pt]
 \V^{(n)}[C,\dots,C]&=&  \int d^4\L_1\cdots d^4\L_n \widetilde V^{(n)}_{\L_1,\dots, \L_n} C_{\L_1}\cdots C_{\L_n}\ ,
\eea
where the symmetric $n$th order plane-waves 
\be \widetilde\Phi^{(n)}_{\L_1,\dots, \L_n}~=~\F^{(n)}[f_{\L_1},\dots,f_{\L_n}]\ ,\qquad  \widetilde V^{(n)}_{\L_1,\dots, \L_n}~=~\V^{(n)}[f_{\L_1},\dots,f_{\L_n}]\ ,\ee 
have well-defined perturbative expansions for open-contour homotopy contractors $\rho$ defined in \eq{collapserho} \cite{Vasiliev:1990vu} (see also \cite{Prokushkin:1998bq})\footnote{It would be interesting to find the exact solution to the Vasiliev equations with initial data $C=g_\L +\pi(g_\L^\dagger )$ where $g_\L=\widetilde C_\L(f_\L+\pi(\tau(f_\L))$ with fixed twistor-space momentum obeying $(\l_\a)^\dagger~=~\bar\l_{\ad}$. In the classical perturbation theory, the twistor plane waves are in the same regularity class as polynomials. In this sense, one may think of the aforementioned solutions as generalizations of the exact $O(3,1)$-invariant instanton solution of \cite{Sezgin:2005pv} (corresponding to $\L=0$). In the language of quantum mechanics, twistor plane waves correspond to coherent states, while squeezed states correspond to ordinary one-particle states. In terms of space-time curvatures, the latter are more localized than the former. This explains why the the perturbative expansion based on initial data corresponding to boundary-to-bulk propagators \cite{Giombi:2009wh,Giombi:2010vg} is more singular than that in the twistor plane-wave sector \cite{Vasiliev:1990vu,Prokushkin:1998bq}. It would be interesting to see whether the closed-contour contractor $\rho_\Gamma$ could be used to simplify the calculations of \cite{Giombi:2009wh,Giombi:2010vg}.}.

\subsubsection{Expansion of master fields up to second order}\label{Expansion of master fields}

Let us compute the first order correction $\widetilde V^{(1)}_{\L_1}=\widehat V^{(1)}[f_{\L_1}]$ and the second order correction $\widetilde \F^{(2)}_{\L_1,\L_2}=\widehat\Phi^{(2)}[f_{\L_1},\L_{2}]$  by integrating \eq{prime} perturbatively keeping the homotopy contours closed. 

\subsubsection*{Calculation of $ \widetilde V^{(1)}_{\L_1}$  }

The first step consists of integrating $q\widetilde V^{(1)}_{\L_{1}} = - f_{\L_1} \star \J={i\over 4}   f_{\L_1} \star (b dz^2\widehat{\kappa}+\bar b d\zb^2\widehat{\kappab})$\,. Using the homotopy operator $\rho_\C$ given in \eqref{j integration} one finds
\be
\widetilde V^{(1)}_{\L_{1}} ~=~ i_Z \oint_{\C_1} {dt_1\over 2\pi i t_1}\gamma(t) \,  t_1^{{\cal L}_Z} \left(  f_{\L_1}  \star {i\over 4}  (b dz^2\widehat{\kappa}+\bar b d\zb^2\widehat{\kappab})\right)
\ee
which is thus given by the holomorphic and anti-holomorphic parts
\be
 f_{\L_1} \star\widehat{\kappa} ~=~  e^{i [ (y + \l_1) z + \yb \lb_1 ]} \qquad  f_{\L_1} \star  \widehat{\kappab} ~ = ~  e^{ i [ + y \l_1  - (\yb+ \lb_1) \zb ]}. 
\ee
Acting with $t_1^{{\cal L}_Z}$ yields
\be
 t_1^{{\cal L}_Z} ~\left( dz^2 ~e^{i [ (y + \l_1) z + \yb\bar{\l}_1 ]} \right) ~=~  t_1^2 ~dz^2 ~e^{i [ (y + \l_1) z t_1 + \yb\bar{\l}_1 ]}
\ee
\be
 t_1^{{\cal L}_Z}~ \left( d\zb^2  ~e^{ i [ + y \l_1  - (\yb+ \lb_1) \zb ]} \right) ~=~t_1^2 d\zb^2 ~e^{ i [ y \l_1  - (\yb+ \lb_1) \zb t_1 ]}\ ,
\ee
which hold for general $t_1\in\C_1$\,.
Let us introduce  an auxiliary twistor momentum $M=(\m,\mb)$ and a source term $e^{iM^{\una}Z_{\una}}$\,, in order to represent each factor $Z^{\una}$ through a partial derivative with respect to $M$\,. The action of the inner product $i_{Z}$ becomes
\bea
i_Z  dz^2 t_1^2 e^{i [ (y + \l_1) z t_1 + \yb\bar{\l}_1 ]}~&=&~ 2  dz^\a z_\a~ t_1^2 ~ e^{i [ (y + \l_1) z t_1 + \yb\bar{\l}_1 ]} \nonumber\\[5pt]
~&=&~ 2  \left.   dz^\a  ~t_1 ~ ( -i)  {\partial\over \partial \m_1^{\a}}     ~ e^{i [ (y + \l_1 + \m_1) z t_1 + \yb\bar{\l}_1 ]}\right|_{\m_1 = 0}
 \eea 
where  $\m_1$ is the first components of the auxiliary twistor momentum $M_1= (\m_1,\mb_1)$ and the derivative operators $ -i  {\partial\over \partial \m_1^{\a}} $ can be treated as a $Z$ independent quantity and factorized.  The expression for the $\zb$ parts similar and one obtains
 \be
 \label{V^1}
\widetilde V^{(1)}_{\L_{1}} ~ =~ {1\over 2}  \left.  \left[ b dz^\a {\partial\over \partial \m_1^{\a}}~+~\bar b  d\zb^\ad{\partial\over \partial \mb_1^{\ad}}\right]  \oint_\C {dt_1\over 2\pi i } ~\gamma(t_1)  ~\left(\phi^+_{\m_1}(\L_1) ~ + ~ \phi^-_{\mb_1}(\L_1)\right)  \right|_{\m_1,\mb_1 = 0}  
\ee
where we have defined
\be\label{v1component} 
\phi^+_{\m_1}(\L_1)~=~  \exp{i [ (y + \l_1 + \m_1) z t_1 + \yb\bar{\l}_1 ]} \qquad  \phi^-_{\mb_1}(\L_1)~=~  \exp{i [y \l_1  - (\yb+ \lb_1 - \mb_1) \zb t_1]}
\ee
that automatically satisfies the gauge condition since $i_Z \widetilde V^{(1)}_{\L_{1}}   \sim z^{\a} z_{\a}t_1 + \zb^{\a} \zb_{\a}t_1 =  0$\,.

\subsubsection*{Calculation of $ \widetilde \Phi^{(2)}_{\L_1,\L_2}$  }
\label{F2}
The next step is the integration of the zero-form equation
\be
\label{order2}
q \, \widetilde \Phi^{(2)}_{\L_1,\L2} ~=~ -\frac12\left([\widetilde V^{(1)}_{\L_1} ,\ f_{\L_2}]_{\pi} ~+~  [\widetilde V^{(1)}_{\L_2} ,\ f_{\L_1}]_{\pi} \right)
\ee
where $\widetilde V^{(1)}_{\L_1}$ is given in ~\eqref{V^1}. The momenta $\L_1, \L_2$ are in general different and $\widetilde \Phi^{(2)}_{\L_1,\L2}$ is symmetrized. The homotopy operation yields
\be \widetilde \Phi^{(2)}_{\L_1,\L2}~=~-\frac12i_Z \oint_{\C_2}{dt_2 \over 2\pi t_2} \gamma(t_2) t_2^{{\cal L}_Z}\left([\widetilde V^{(1)}_{\L_1} ,\ f_{\L_2}]_{\pi} ~+~  [\widetilde V^{(1)}_{\L_2} ,\ f_{\L_1}]_{\pi} \right)
\ .\label{order2b}\ee
We split  the first $\star$-commutators on the  r.h.s. of ~\eqref{order2} in two parts following the definitions in \eqref{v1component} and perform the $\star$-products using     
\bea
\label{formula1}
 f_{\L}~\star~\widehat f(y,\yb;z,\zb)&=&f_{\L}~\widehat f (y + \l, \yb+ \lb;z- \l , \zb+ \lb)\ ,\\[5pt]  \widehat f(y,\yb;z,\zb)~\star~f_{\L}&=&f_{\L}~\widehat f(y - \l, \yb - \lb ; z - \l , \zb+ \lb)\ ,
\eea
to  arrive at the following contribution:
\be\label{pi-commutator+} 
-\frac 12 [ \phi^+_{\m_1}(\L_1) , \ f_{\L_2} ]_{\pi}   ~=~ e^{ i\left[ y\l_2+\yb \bar{\l}_2 + y (z - \l_2) t_1+\yb \bar{\l}_1 \right] } \cos{\left[ \lb_2 \bar{\l}_1 -(\l_1- \l_2 + \m_1)(z - \l_2) t_1\right] }\ ,
\ee 
and a similar expression for the complex conjugate. To act with $  i_Z t_2^{{\cal L}_Z}$ on the one-forms in \eqref{order2b} we introduce another auxiliary twistor momentum $M_2=(\mu_2,\bar \mu_2)$ with its source term $e^{i M_2^{\una} Z_{\una}}$\,, such that 
\be
  i_Z t_2^{{\cal L}_Z}   \left( dz^\a \partial_{\m_1^{\a}}   [ \phi^+_{\m_1}(\L_1)  , \ f_{\L_2} ]_{\pi}\right)~=~  \left. \partial_{\m_1} \partial_{ \m_2} \left( e^{i ~\m_2 z \,t_2}   t_2^{{\cal L}_Z} [ \phi^+_{\m_1}(\L_1) , \ f_{\L_2} ]_{\pi}\right)\right|_{\m_1=\m_2=0}
 \ee
\emph{idem.} the $\phi^-_{\mb_1}(\L_1)$-contribution, where we have used $\left(-i {\partial\over \partial \m_2^{\b}} \right)  \left(-i {\partial\over \partial \m_1^{\a}} \right) \e^{\a\b}=\partial_{\m_1} \partial_{ \m_2}$ with $\partial_{\m_1} \partial_{ \m_2}=\partial^\a_{\m_1} \partial_{ \m_2,\a}$\,.
Substituting the explicit form of  $ - [ \phi^+_{\m_1}(\L_1) , \ f_{\L_2} ]_{\pi}$ given in \eqref{pi-commutator+} and $ - [ \phi^-_{\mb_1}(\L_1) , \ f_{\L_2} ]_{\bar \pi}$, we find
 \bea
 \label{F2expression}
 \widetilde \Phi^{(2)}_{\L_1\L_2}& = &\frac{i}2 \left[  b\partial_{\m_1} \partial_{ \m_2}   + \bar b \partial_{\mb_1} \partial_ {\mb_2}  \right]~\left(\prod_{i=1,2}\oint_{\C_i} {dt_i \over 2\pi i t_i } \gamma(t_i)\right)\\[5pt] &&\left.\times~ t_1 \left\{ \phi^+_{\m_1,\m_2} (\L_1,\L_2) + \phi^-_{\mb_1,\mb_2}(\L_1,\L_2)+ (\L_1\leftrightarrow \L_2)\right\} \right|_{M_1,M_2=0}\ ,\nonumber
 \eea
where we have defined
\bea
\phi^+_{\m_1\m_2}( \L_1,\L_2)  &=& e^{ i \left[ y \l_2  (1 - t_1) + \yb (\lb_1 + \lb_2 ) + (y + {\m_2\over t_1}) z~t_1 t_2 \right]} ~ \cos{\left[ \lb_2 \bar{\l}_1 - (\l_1- \l_2 + \m_1)(z~t_2 - \l_2) t_1\right]}\ ,\qquad 
\eea
and its the complex conjugate  $\phi^-_{\mb_1\mb_2}( \L_1,\L_2)$. So far, the choice of closed contours is immaterial, since no divergences have appeared. Thus, the closed contours could in principle be converted into open line integrals. However, as we shall see next, the closed contour prescription will be crucial in calculating invariant quantities.

\subsubsection{Evaluation of zero-form charges in the leading order}
 
Let us evaluate the quasi-amplitudes corresponding to the leading orders \eq{I1}--\eq{I3} of the zero-form charges.
Using 
\be
\label{effective product}
 f_{\L_1} \star\pi( f_{\L_2}) \star  f_{\l_3} \star \cdots \star f_{\L_{\NumbCurv-1}} \star \pi( f_{\L_{\NumbCurv}}) ~=~   e^{i[\Th_{1\cdots \NumbCurv} + \Thb_{1\cdots \NumbCurv}] } f_{\L_{1\cdots \NumbCurv}}
\ee
where  
\be
\label{effective momentum}
\l_{1\dots n}~=~ \sum_{i=1}^{n} (-)^{i-1} \l_{i},  \quad~ \lb_{1\dots (n)}~=~ \sum_{i=1}^{n}  \lb_{i}\ ,
\ee
\be
\label{effective classical}
\Th_{1\dots (n)}~=~ \sum_{i > j }^{n}  \l_{i} \l_{j},  \quad~ \Thb_{1\cdots (n)}~=~ - \sum_{i>j}^{n}  \lb_{i}\lb_{j}\ ,\ee
one find the following quasi-amplitudes in the leading order
\be \widetilde \cI^{(K)}_K(\L_1,\dots,\L_K) ~=~
  \frac{1}{\NumbCurv!} \sum_{perm.}~ e^{i[\Th_{  1\dots \NumbCurv  } + \Thb_{  1\dots \NumbCurv  } ]  }\ ,
\ee
\be (\widetilde \cI^++\widetilde \cI^-)^{(K+1)}_{K+1}(\L_1,\dots,\L_{K+1}) ~=~
  \frac{1}{(\NumbCurv+1)!} \sum_{perm.}~ 2\pi \,\delta^2(\l_{1\dots (K+1)})\, e^{i[\Th_{  1\dots (\NumbCurv+1)  } + \Thb_{  1\dots (\NumbCurv +1) } ]  }\ ,
\ee

\be \widetilde \cI^{\prime(K)}_K(\L_1,\dots,\L_n) ~=~
  \frac{1}{\NumbCurv!} \sum_{perm.}~ (2\pi)^2\, \delta^2(\l_{1\dots K})\,\delta^2(\bar\l_{1\dots K})\,e^{i[\Th_{  1\dots \NumbCurv  } + \Thb_{  1\dots \NumbCurv  } ]  }\ .
\ee

\subsubsection{Evaluation of zero-form charges at next-to-leading order }
\label{Quasi-amplitudes: a first look at sub-leading orders}

In what follows we use the large-contour prescription to evaluate the first sub-leading terms in the expansion of the zero-form charges \eq{Inv1}--\eq{Inv3} in the initial data $C$ defined in \eq{defC}. We shall assume that the gauge function drops out from the trace, so that $d\cI_{\NumbCurv}=0$ holds manifestly and we can work at the point where $X=P=0$. For simplicity of notation, we shall drop the primes in the remainder of this section.

\subsubsection*{${\cal I}_K$ at next-to-leading order }

We first treat the zero-form charges $ \cI_{\NumbCurv}$ defined in \eqref{Inv1}, its perturbative expansion reads $(\NumbCurv=2,4,6,\dots$)
\be
\label{perturbative invariant} \cI_{\NumbCurv}~=~\sum_{n=0}^\infty \cI_{\NumbCurv}^{(\NumbCurv+n)}\ ,\qquad 
\cI_{\NumbCurv}^{(\NumbCurv + n)}~=~ \frac{1}{(\NumbCurv + n)!}\sum_{\tiny\ba{c} n_1+\dots+n_{\NumbCurv}\\ = \NumbCurv + n\ea }\widehat{Tr}\left[ \widehat\Phi^{(n_1)}\star\dots\star \pi(\widehat\Phi^{(n_{\NumbCurv})}) \star \widehat{\kappa} \widehat{\bar{\kappa}}\right] \ .
\ee
In the plane-wave basis, the corresponding quasi-amplitudes read\footnote{The sum over permutations exchanges the $\NumbCurv+n$ external twistor momenta. Before explicit symmetrization the separate contributions $\widetilde\cI_\NumbCurv^{(\NumbCurv+n)}$ to the quasi-amplitudes have partial symmetries due to the cyclic property of the trace operation and the total symmetry of the higher-order plane waves.}
\be  \label{invariant in plane waves}
\widetilde \cI_{\NumbCurv}^{(\NumbCurv+n)}(\L_1,\dots,\L_{\NumbCurv+n})~=~  \sum_{\tiny\ba{c} n_1+\dots+n_{\NumbCurv}\\ = \NumbCurv + n\ea }  \widetilde\cI_\NumbCurv^{(n_1,\dots,n_\NumbCurv)}(\L_1,\dots,\L_{\NumbCurv+n})\ ,\ee\be 
\widetilde\cI_\NumbCurv^{(n_1,\dots,n_\NumbCurv)}(\L_1,\dots,\L_{\NumbCurv+n})~=~
\frac{1}{(\NumbCurv + n)!} \sum_{perm} 
\widehat{Tr}
\left[ \widetilde\Phi^{(n_1)}_{\unL_{1}^{(n_1)}}\star\dots\star \pi(\widetilde\Phi^{(n_{\NumbCurv})}_{\unL^{(n_\NumbCurv)}_{{\NumbCurv}}})\star \widehat{\kappa} \widehat{\bar{\kappa}}\right]  \ ,\ee
where $\unL^{(n)}~=~ (\L_{1}, \L_{2}\cdots\L_{n} )$ and we the denote the $n$th order plane waves by  
\be \widetilde \Phi^{(n)}_{\unL^{(n)}}~=~\widetilde \Phi^{(n)}_{\L_1\dots\L_n}~=~\widetilde \Phi^{(n)}_{\L_1\dots\L_n} (f_{\L_1}, \dots f_{\L_n} )\ ,\ee 

The first sub-leading correction $\widetilde \cI^{(\NumbCurv+1)}_\NumbCurv=\frac{\NumbCurv}2 \left(\cI^{( 2,1,\dots,1)}_\NumbCurv+\cI^{( 1,2,1\dots,1)}_\NumbCurv\right)$ contain formally singular integrals over $Y$ and $Z$ space. As we shall demonstrate next, choosing to regularize these  the large-contour scheme yields vanishing next-to-leading orders ,  \be \widetilde \cI^{(\NumbCurv+1)}_\NumbCurv~=~0\ ,\ee
where the zeroes are of the form 
\be \left[\oint_{\Gamma_1}{dt_1\over 2\pi i t_1}
\oint_{\Gamma_2}{dt_1\over 2\pi i t_2} \gamma(t_1) \, \frac{1}{(1-t_1 t_2)^2}\right]\times \l^2 \ ,\ee
that is, a potential logarithmic divergence times a vanishing square of an external twistor momentum $\l$.

To this end, let us first consider the case $\NumbCurv=2$\,.  
We replace one of the linearized plane waves with the second order plane-wave $\widetilde \Phi^{(2)}_{\L_1\L_2}$ given in \eqref{F2expression}, that is
\be
\label{I2correction}
\widetilde \cI^{(1,2)}_{2}~=\frac{1}{3!}~\widehat{Tr} \left[ \left(\widetilde \Phi_{\L_1\L_2}^{(2)}\star\pi(f_{\L_3}) ~+~ f_{\L_1} \star\pi(\widetilde \Phi_{\L_1\L_2}^{(2)}) \right)\star \widehat{\kappa} \widehat{\bar{\kappa}}\right] 
\ee
The second-order plane wave is the sum of two contributions $\phi^+_{\m_1\m_2}( \L_1,\L_2)$  and $\phi^-_{\m_1\m_2}( \L_1,\L_2)$ that we can treat separately.  We obtain four different contributions, namely
\bea\label{contributions+}
f_{\L_3} \star \pi ( \phi^+_{\m_1\m_2}( \L_1,\L_2))\star \widehat{\kappa} \widehat{\bar{\kappa}}&=&
e^{i (yz - \yb\zb -z \l_3 + \zb \lb_3 )} e^{i [ (z-\l_3) \l_2  (1 - t_1) + (\zb+\lb_3) (\lb_1 + \lb_2 ) + [(z-\l_3) + {\m_2\over t_1}] (y + \l_3)~t_1 t_2 ]} \nonumber\\[5pt]&&\times ~
\cos{\left[ \lb_2 \lb_1 - (\l_1- \l_2 + \m_1)[(y + \l_3)~t_2 - \l_2] t_1\right]}\ ,\\[5pt]
\phi^+_{\m_1\m_2}( \L_1,\L_2) \star \pi (f_{\L_3})\star \widehat{\kappa} \widehat{\bar{\kappa}}&=&
e^{i (yz - \yb\zb + z \l_3 + \zb \lb_3 )}e^{i [ - (z-\l_3) \l_2  (1 - t_1) + (\zb - \lb_3) (\lb_1 + \lb_2 ) + [ (z- \l_3) - {\m_2\over t_1}] (y - \l_3)~t_1 t_2 ]} \nonumber\\[5pt]&&\times ~
\cos{\left[ \lb_2 \lb_1 + (\l_1- \l_2 + \m_1)[(y - \l_3)~t_2 - \l_2] t_1\right]}\ ,
\eea
where we have applied \eqref{formula1}, and two related expressions for $f_{\L_3}\star \pi (\phi^-_{\mb_1\mb_2}( \L_1,\L_2))  \star \widehat{\kappa} \widehat{\bar{\kappa}}$ and $\phi^-_{\mb_1\mb_2}( \L_1,\L_2) \star \pi (f_{\L_3})\star \widehat{\kappa} \widehat{\bar{\kappa}}$\,. In the case of $\phi^+$-contributions, integration over $d^2z$ and $d^2\zb$ gives
\bea\label{delta-uno}
\int d^2z e^{i z [ -( y + \l_3) + \l_2 (1 -  t_1) + (y + \l_3) ~t_1 t_2 ]} ~=~ 2\pi~ \d^2 \left( (-y -\l_3) (1- t_1 t_2) + \l_2 (1 - t_1)\right) \\
\qquad = \frac{ 2\pi}{(1 - t_1 t_2)^2} \d^2 \left( y + \l_3 - \l_2 \frac{(1 - t_1)}{1-  t_1 t_2}\right) \nonumber 
\eea
\be
\int d^2 \zb e^{i \zb [ -\yb +\lb_3 + \lb_2 + \lb_1]}   ~=~ 2\pi~  \d^2 \left( \yb +\lb_3 + \lb_2 + \lb_1 \right) \ ,
\ee
and likewise
\bea
\label{delta-due}
\int d^2z e^{i z [ -y +\l_3 - \l_2 (1- t_1) + ( y+ \l_3)  ~t_1 t_2 ]} ~=~ 2\pi~ \d^2 \left( (-y +\l_3) (1- t_1 t_2) - \l_2 (1 - t_1)\right) \\
\quad = \frac{ 2\pi}{(1 - t_1 t_2)^2} \d \left( y - \l_3 + \l_2 \frac{(1 - t_1)}{1-  t_1 t_2}\right)\nonumber
\eea
\be
\int d^2\zb e^{i \zb [ \yb +\lb_3 + \lb_2 + \lb_1]}  ~=~ 2\pi~\d^2 \left( -\yb +\lb_3 + \lb_2 + \lb_1 \right) \ ,
\ee
where the $y$-space $\d^2$-functions in \eqref{delta-uno} and \eqref{delta-due} contain the potentially divergent factor $\frac{ 2\pi}{(1 - t_1 t_2)^2}$ depending on the homotopy integration variables that must be regularized using the large-contour prescription. In fact, when we push both the $t_1$ and $t_2$ contours to infinity  the parameters $t_1$ and $t_2$ are large  everywhere and  $(1 - t_1 t_2)^2 \neq 0$\,.  This makes the  $\d^2$-functions well-defined and we can safely perform the $Y$-space integration, \emph{viz.} 
\bea
\widehat{Tr}[f_{\L_3} \star\pi(\phi^+_{\m_1\m_2}( \L_1,\L_2))\star \widehat{\kappa} \widehat{\bar{\kappa}} ] ~=~ \frac{ 2\pi}{(1 - t_1 t_2)^2}  e^{i [ +\lb_3(\lb_1 + \lb_2 ) +\l_2 \l_3 [ \frac{(1 - t_1) t_1 t_2}{1-  t_1 t_2} - (1 - t_1)] +   \m_2 \l_2 \frac{(1 - t_1) t_2}{1-  t_1 t_2} ]} \nonumber \\
\qquad\cos{[ \lb_2 \lb_1 - (\l_1- \l_2 + \m_1)(\l_2\frac{(1 - t_1) t_1 t_2}{1-  t_1 t_2}  + \l_2 t_1)]}\ ,\qquad
\eea
\bea
\widehat{Tr}[\phi^+_{\m_1\m_2}( \L_1,\L_2) \star\pi(f_{\L_3})\star \widehat{\kappa} \widehat{\bar{\kappa}}] ~=~  \frac{ 2\pi}{(1 - t_1 t_2)^2}  e^{i [  - \lb_3(\lb_1 + \lb_2 )  - \l_2 \l_3 [ \frac{(1 - t_1) t_1 t_2}{1-  t_1 t_2} + (1-t_1) ] -   \m_2 \l_2 \frac{(1 - t_1) t_2}{1-  t_1 t_2} ]} \nonumber \\
\qquad\cos{[ \lb_2 \lb_1 + (\l_1- \l_2 + \m_1)(\l_2\frac{(1 - t_1) t_1 t_2}{1-  t_1 t_2}  -  \l_2 t_1)]}\ .
\eea
Now we observe that the two auxiliary momenta $\m_1$ and $\m_2$ always appear contracted into the same external twistor momentum $\l_2$\,. Hence, when acting with the differential operator $\partial_{\m_1}\partial_{\m_2}$ as in \eqref{F2expression}, the above two  contributions vanish, \emph{viz.}
\be
\partial_{\m_1}\partial_{\m_2}  \widehat{Tr}[f_{\L_3} \star\pi(\phi^+_{\m_1\m_2}( \L_1,\L_2))\star \widehat{\kappa} \widehat{\bar{\kappa}}] ~\sim~ \l_2 \l_2~ = ~0
\ee
and 
\be
\partial_{\m_1}\partial_{\m_2}   \widehat{Tr}[\phi^+_{\m_1\m_2}( \L_1,\L_2) \star\pi(f_{\L_3} )\star \widehat{\kappa} \widehat{\bar{\kappa}}] ~\sim~  \l_2 \l_2~ =~ 0\ .
\ee
A related cancellation occurs for the $\phi^{(-)} $-contributions, and we conclude that\be \widetilde \cI^{(3)}_{2}(\L_1,\L_2,\L_3)~=~\frac{1}{3!}\widehat{Tr}[(f_{\L_3} \star\pi(\widetilde \Phi^{(2)}_{\L_1,\L_2}) + \widetilde \Phi^{(2)}_{\L_1,\L_2}\star\pi(f_{\L_3}) )\star \widehat{\kappa} \widehat{\bar{\kappa}}]  ~=~ 0\ . \ee 
This result generalizes immediately to any $\NumbCurv$\,, where one finds that $\cI^{(\NumbCurv+1)}_\NumbCurv$ are given by vanishing contributions of the form  
$\widehat{Tr}[f_{\L_{1\dots (\NumbCurv-1)}} \star \pi( \widetilde \Phi_{\L_{\NumbCurv}\L_{\NumbCurv+1}}^{(2)})\star\widehat\kappa\widehat{\bar\kappa}]$ and  $\widehat{Tr}[\pi( f_{\L_{1\dots (\NumbCurv-1)}}) \star  \widetilde \Phi_{\L_{\NumbCurv}\L_{\NumbCurv+1}}^{(2)}\star\widehat\kappa\widehat{\bar\kappa}]$ with total twistor momentum given by \eq{effective momentum}.

\subsubsection*{${\cal I}^\pm_{K+1}$ at next-to-leading order}
\label{odd invariants}

We next turn to the zero-form charges ${\cal I}^\pm_{K+1}$ defined in \eq{Inv2}. 
The first sub-leading term in ${\cal I}^\pm_{3}$ is obtained by replacing in  $\widetilde{\cal I}_{3}^{\pm (3)}(f_{\L_1}, f_{\L_2},f_{\L_3}) $ one of the linearized plane waves with the second order plane-wave $\widetilde \Phi^{(2)}_{\L_1\L_2}$, given in \eqref{F2expression}, that is
\bea \label{I_3^4}
\widetilde{\cal I}_3^{\pm (4)} (\L_1,\L_2,\L_3,\L_4)~&=&~\frac1{(3+1)!}\sum_{\tiny\mbox{permutations}} \frac12(Tr_y STr_{\bar y}\pm STr_y Tr_{\bar y})\\
&& \hspace{-1cm}\times \left[  \widetilde \Phi^{(2)}_{\L_1\L_2} \star \pi(f_{\L_3})\star f_{\L_4}~+~ f_{\L_1} \star \pi(\widetilde \Phi^{(2)}_{\L_2\L_3})\star f_{\L_4} ~+~f_{\L_1} \star \pi(f_{\L_2})\star \widetilde \Phi^{(2)}_{\L_3\L_4}\right]\ \nonumber  ,
\eea
We want to show the vanishing of all the relevant contributions to $\widetilde{\cal I}_3^{\pm (4)} (\L_1,\L_2,\L_3,\L_4)$ coming from these three terms vanish, as in the $K=2$ case considered above. Here we have to apply the composite trace operator $ \frac12(Tr_y STr_{\bar y}\pm STr_y Tr_{\bar y}) $ corresponding to a simple trace and the insertion of a $\star  \frac{ \widehat\kappa \pm \widehat{\bar\kappa}}{2}$ factor. We can again treat separately the holomorphic and anti-holomorphic part of $\widetilde \Phi^{(2)}_{\L\L}$ and show that for the $\phi^+_{\m_1\m_2}( \L_1,\L_2)$ part we have  
\be \label{example}
Tr \left[ \phi^+_{\m_1\m_2}( \L_1,\L_2)  \star \pi(f_{\L_3})\star f_{\L_4} \star  \frac{ \widehat\kappa \pm \widehat{\bar\kappa}}{2} \right] ~\sim~ \l^2 ~\pm~ {\cal N} ~\text{contributions} ~\sim~ 0
\ee
\be 
Tr \left[ f_{\L_1}~ \star~ \pi( \phi^+_{\m_1\m_2}( \L_2,\L_3) )~\star ~f_{\L_4} ~\star  ~ \frac{ \widehat\kappa \pm \widehat{\bar\kappa}}{2} \right] ~\sim~ \l^2 ~\pm~ {\cal N} ~\text{contributions} ~\sim~ 0~
\ee
\be 
Tr \left[ f_{\L_1} ~\star ~\pi( f_{\L_2} )~\star  ~\phi^+_{\m_1\m_2}( \L_3,\L_4)~ \star ~ \frac{ \widehat\kappa \pm \widehat{\bar\kappa}}{2} \right] ~\sim~ \l^2 ~\pm~ {\cal N} ~\text{contributions} ~\sim~ 0~
\ee
where  $ {\cal N}^{-1}~=~\int \frac{d^2z}{2\pi} $  (see \eq{IN}).  
Let us show how this happens. The first and the third contribution are similar and can be expressed using an effective momentum using \eq{formula1}, \emph{i.e.} 
\be\label{uno}
\widetilde \Phi^{(2)}_{\L_1\L_2}~ \star ~\pi(f_{\L_3})\star f_{\L_4}~=~ e^{i\Theta_{34}}~  \widetilde \Phi^{(2)}_{\L_1\L_2} ~\star~ f_{\L_{34}}~=~ e^{i\Theta_{34}}~ f_{\L_{34}} ~ \widetilde \Phi^{(2)}_{\L_1\L_2} (y-\l_{34}, \yb-\lb_{34}; z-\l_{34}, \zb +\lb_{34}) 
\ee
\be\label{due}
  f_{\L_1}\star~ \pi(f_{\L_2})\star~ \widetilde \Phi^{(2)}_{\L_3\L_4} ~=~ e^{i\Theta_{12}} ~ f_{\L_{12}} ~\star~ \widetilde \Phi^{(2)}_{\L_3\L_4} ~=~ e^{i\Theta_{12}} ~ f_{\L_{12}}~   \widetilde \Phi^{(2)}_{\L_3\L_4} (y+\l_{12}, \yb+\lb_{12}; z-\l_{12}, \zb -\lb_{12}) 
\ee
where we have defined the total momenta $ \L_{34} = (\l_3-\l_4, \lb_3+\lb_4 )$, $ \L_{12} = (\l_1-\l_2, \lb_1+\lb_2 )$ and the classical parts $ \Theta_{34} = \l_4\l_3 - \lb_4\lb_3 $ $\Theta_{12} = \l_2\l_1, - \lb_2\lb_1 $.  The second one should be considered apart and we obtain   
\be\label{tre}
f_{\L_1} \star \pi(\widetilde \Phi^{(2)}_{\L_2\L_3}  )~\star~ f_{\L_4}~=~ e^{i\Theta_{14}}  ~f_{\L_{14}} ~ \widetilde \Phi^{(2)}_{\L_2\L_3} (- y-\l_1+ \l_4, \yb-\lb_1+ \lb_4; - z +\l_1+ \l_4, \zb + \l_1+ \l_4) \, 
\ee
where $ \L_{14} = (\l_1+ \l_4, \lb_1 + \lb_4 )$ and $ \Theta_{14} = - \l_4\l_1 - \lb_4\lb_1 $. 
We start by considering the first contribution in  \eq{example}, that turns out to be the sum of two terms, proportional to $  \widehat\kappa $ and $ \widehat{\bar\kappa}$ respectively. The $ \widehat\kappa$-part is given by  
\be
Tr \left[ \phi^+_{\m_1\m_2}( \L_1,\L_2)  \star \pi(f_{\L_3})\star f_{\L_4} \star  \widehat \kappa  \right] ~=~ \nonumber\ee
\be \frac{2 \pi~ e^{i \Theta_{34}}}{(1- t_1t_2)^2}  \int d^2\zb  ~e^{i \left[ -\l_{34} \l_2 (1-t_1) - \lb_{34}(\lb_1+\lb_2) + (- \l_{34} - \frac{\m_2}{t_1} ) \frac{\l_2(1-t_1) t_1 t_2}{1-t_1t_2}t_1t_2 \right]} \nonumber \ee\be
 \times \,\cos{\left[ \lb_2 \lb_1 - (\l_1- \l_2 + \m_1) (\frac{\l_2(1-t_1) t_1 t_2}{1-t_1t_2}t_1t_2 - \l_2 t_1) \right]} \,
\ee
which vanishes when we act with two $\mu_1\mu_2$ contracted derivatives 
\be
\partial_{\m_1} \partial_{ \m_2} ~Tr \left[ \phi^+_{\m_1\m_2}( \L_1,\L_2)  \star \pi(f_{\L_3})\star f_{\L_4} \star  \widehat \kappa  \right]  ~\sim ~\l_2^2 ~=~ 0
\ee
Instead, the insertion of $ \widehat{\bar\kappa}$ produce a non-vanishing contribution, \emph{viz.}
\be 
Tr \left[ \phi^+_{\m_1\m_2}( \L_1,\L_2)  \star \pi(f_{\L_3})\star f_{\L_4} \star  \widehat{\bar\kappa}  \right] ~=~\ee\be  \frac{2\pi~e^{i \Theta_{34}}}{(t_1t_2)^2} ~\int d^2 \ \yb \ d^2 z~ ~\delta^2(z- \l_{34} + \frac{\l_{34} + \l_2 (1-t_1)}{t_1t_2}) \,\delta^2(\lb_1+\lb_2+\lb_{34} - \yb ) \nonumber\ee\be\times \,e^{i \left[   -\l_{34} \l_2 (1-t_1) - \lb_{34}(\lb_1+\lb_2)  -  ( \l_{34} + \frac{\m_2}{t_1} ) (z- \l_{34}) t_1t_2 \right]}\,  \cos{\left[ \lb_2 \lb_1 - (\l_1- \l_2 + \m_1) \left( (z- \l_{34}) t_1t_2 - \l_2 t_1\right) \right]} 
 \ee
 but it does not contain the infinite chiral volume $\int d^2\zb$ and is suppressed by the normalization infinite factor ${\cal N}$. 
In the anti-holomorphic part $Tr \left[ \phi^-_{\mb_1\mb_2}( \L_1,\L_2)  \star \pi(f_{\L_3})\star f_{\L_4} \star \frac{ \widehat\kappa \pm \widehat{\bar\kappa}}{2}\right] $ the r\^ole of $ \widehat\kappa$ and $ \widehat{\bar\kappa}$ is exchanged and we find     
\be
\partial_{\mb_1} \partial_{ \mb_2} Tr \left[ \phi^-_{\mb_1\mb_2}( \L_1,\L_2)  \star \pi(f_{\L_3})\star f_{\L_4} \star \frac{ \widehat\kappa \pm \widehat{\bar\kappa}}{2} \right] ~\sim~  {\cal N}~\text{contributions}~  \pm ~ \lb^2 ~\sim~ 0
\ee
\be
\partial_{\mb_1} \partial_{ \mb_2} Tr \left[ f_{\L_1} \star \pi( \phi^-_{\mb_1\mb_2}( \L_2,\L_3) )\star f_{\L_4}\star \frac{ \widehat\kappa \pm \widehat{\bar\kappa}}{2} \right] ~\sim~  {\cal N} ~\text{contributions}~ \pm ~ \lb^2 ~\sim~ 0
\ee
\be
\partial_{\mb_1} \partial_{ \mb_2} Tr \left[ f_{\L_1} \star \pi( f_{\L_2} )\star  \phi^-_{\mb_1\mb_2}( \L_3,\L_4)\star \frac{ \widehat\kappa \pm \widehat{\bar\kappa}}{2} \right] ~\sim~  {\cal N} ~\text{contributions}~ \pm ~ \lb^2 ~\sim~ 0
\ee
that allows us to conclude that 
\be
\widetilde{\cal I}_3^{\pm (4)} (\L_1,\L_2,\L_3,\L_4) ~= ~0
\ee
Then, using the total momentum technique that we have explained in the $\cI^{(N+1)}_{N}$ computation, toghether with the cyclic property of the trace, we can generalize the result to the general case, \emph{i.e.}
\be
\widetilde{\cal I}_{N+1}^{\pm ( N+2)} (\L_1,\dots,\L_{N+2}) ~= ~0
\ee

\subsubsection*{${\cal I}^{\prime}_{K}$ at next-to-leading order}

An argument similar argument to the one used above explains that the next-to-leading order terms vanish also in the case of the zero-form charges $\cI'_N$ defined in \eq{Inv3}. In other words, these corrections are canceled by the normalization factor ${\cal N}^2$, and one has
\be
\widetilde{\cal I'}_N^{(N+1)} (\L_1\dots,\L_{N+1}) ~=~0\ .
\ee

\subsection{Discussion: transgression and all-order protection?}\label{Transgression}

On may ask whether it could not be the case that all sub-leading corrections vanish identically in the plane-wave sector. In general, looking at more complicated sectors, the first sub-leading correction $\cI^{(3)}_2$ is a building block for three-point functions which are usually constrained by symmetry. However, the cancellations replicate for arbitrary $\NumbCurv$ which begs for an explanation. 
By analogy with Yang-Mills theory one may think about the integration over $Y$-space as tracing over various representations in the fiber, while the integration over $Z$-space plays the r\^ole of integration in the base manifold, \emph{viz.} 
\be {\cal I}_{\NumbCurv}~=~ -8\,\widehat{Tr} [\widehat F\star \widehat F\star \widehat \Psi^{\star(\NumbCurv-2)}]\label{Chern}\ ,\ee 
where we use the notation given in \eq{defPsi} and it is understood that the trace operation peels off the volume form $d^2z d^2\bar z$ in $Z$-space.
Thus, for $\NumbCurv=1$ one thus has the standard transgression formula
\be {\cal I}_2~=~ -8\,\widehat{Tr}\left[ q\left(\V\star q\V+\frac23 \V^{\star 3}\right)\right]\ ,\ee 
where the exterior derivative has been pulled out without making use of any homotopy contraction operator, which suggests that $\cI_2^{(n)}$ vanish for all $n\geqslant 3$\,.
This leads to the natural question whether it is possible to pull out a $q$ from the integrand in \eq{Chern} without using any homotopy operator, that is, by using only the constraints, which can be decomposed into  holomorphic and anti-holomorphic components as follows:
\be  \partial  ~\widehat U + \widehat U\star  \widehat U~=~ \Omega~ \widehat \Psi\ ,\qquad
\bar\partial  ~\widehat {\overline U} + \widehat {\overline U}\star  \widehat {\overline U}~=~ \overline \Omega~ \widehat {\overline \Psi}\ ,\ee 
\be  \partial ~\widehat {\overline U}+\bar \partial \widehat { U}+ \widehat { U}
\star \widehat {\overline U}+\widehat {\overline U}\star \widehat { U}~=~0\ ,\ee 
\be \bar\partial ~\widehat \Psi+ \widehat {\overline U}\star \widehat\Psi-\widehat\Psi\star
\widehat{\overline U}~=~0\ ,\qquad 
\partial ~\widehat {\overline\Psi}+ \widehat {U}\star \widehat{\overline\Psi}-\widehat{\overline\Psi}\star
\widehat{ U}~=~0\ ,\ee 
where we have defined 
\be  \partial~=~dz^\a \partial_\a\ ,\quad \bar \partial~=~d\zb^\ad \partial_\ad\ , \qquad \widehat U~=~dz^\a\widehat V_\a \ ,\quad \widehat{\overline U}~=~d\zb^\ad \V_{\ad}\ , \ee 
\be  \Omega~=~\frac{ib}4 dz^2\ ,\qquad \overline \O~=~ -(\Omega)^\dagger~=~ \frac{i\bar b}4 d\zb^2\ .\ee 
Indeed, in the next-to-leading order in perturbation theory, the following holomorphic transgression formula holds for all $\NumbCurv$:
\be \cI^{(\NumbCurv+1)}_\NumbCurv= \ft{-32 \NumbCurv}{\NumbCurv+1}\,\widehat{Tr}\left[\bar\partial\left\{\Omega\left[\widehat{\overline U}^{(2)}  (\widehat\Psi^{(1)})^{\NumbCurv-1}+\frac12 \widehat U^{(1)}\left(\widehat\Psi^{(2)}(\widehat\Psi^{(1)})^{\NumbCurv-2}+\cdots+(\widehat\Psi^{(1)})^{\NumbCurv-2}\widehat\Psi^{(2)}\right)\right]\right\}\right]\ ,\ee
where we have suppressed the stars, but we have not found any generalization to higher orders. 

\section{Conclusions}

In this paper we have addressed an aspect of the issue of localizability in Vasiliev's four-dimensional higher-spin gravity: the aim is the identification of sub-sectors of the classical moduli space consisting of solutions with well-defined centers of mass that exhibit cluster-decomposition at the level of a suitable set of classical observables. 
More precisely, we have focused on the regularization of a particular set of classical observables, referred to as zero-form charges, that do not break any higher-spin gauge symmetries and that depend only on the locally defined curvatures and their derivatives in a single coordinate chart. These observables can be defined non-perturbatively and then be given a double perturbative expansion: one first expands such a charge, $\cI$ say, in the Weyl zero-form comprising all curvatures and their derivatives on shell; 
at each order, $n$ say, this yields an $n$-linear gauge-equivariant functional $\cI^{(n)}$ on the infinite-dimensional Weyl zero-form module; one then fixes a specific representation of the Weyl zero-form module, which essentially amounts to choosing boundary conditions, after which $\cI^{(n)}$ can be given a separate expansion in terms of the quantum numbers used to label the states of the representation in question and interpreted as the basic building blocks for dual amplitudes that we refer to as quasi-amplitudes.

The zero-form charges are functionals $\cI[\F]$ of the full Weyl zero-form $\F$ given by are integrals over the twistor $(Y,Z)$-space of star polynomials in $\widehat \Phi'$ and $\pi(\widehat \Phi')$ where $\F':=\F|_{p}$ is the value of $\F$ at a single point $p\in M$. This value can in its turn be expanded perturbatively in terms of the initial datum $C:=\F'|_{Z=0}$, introducing auxiliary contour integrals used to homotopy contract the exterior derivative on the twistor $Z$-space. The resulting perturbative $C$-expansions of the locally accessible zero-form charges $\cI[\F]$  are thus given by nested auxiliary and twistor-space integrations, where the latter come from star products and the trace operation. 

Depending on the choice external states in the corresponding quasi-amplitudes, these nested integrals exhibit various singularity structures: 
one type stems from strongly coupled derivative expansions in spacetime and presents itself already at the level of the perturbative expansion of the locally defined fields (prior to inserting them into the zero-form charges); 
another type stems from tracing over the doubled twistor space.
Physically speaking, the former divergencies are important for the ordinary holographic approach based on spacetime lowest-weight states that can be described perturbatively by boundary-to-bulk propagators: 
indeed the authors of \cite{Giombi:2009wh,Giombi:2010vg} have found that there exist deformed integration contours in twistor space turning such potential divergencies into well-defined residues reproducing three-point correlation functions of free-field currents in accordance with the proposals of \cite{Sundborg:2000wp,Sezgin:2002rt,Klebanov:2002ja}.

In this paper we have instead examined the second type of divergencies that appear already in the sector of twistor-space plane waves (that has been proposed to correspond to amplitudes of topological open strings \cite{Engquist:2005yt}). We have proposed a regularization scheme based on keeping the twistor-space contours fixed and instead performing the auxiliary integrations over large closed contours (see Sections \ref{Expansion in the Weyl zero-form} and \ref{Master fields: perturbative associativity}). This prescription respects associativity and hence higher-spin gauge invariance for external states such that no poles appear in between the auxiliary closed contours in the limit when they are all taken to be large enough (which essentially amounts to that the regularized values are given by the collection of residues at the infinities in the complex planes of the auxiliary integration variables). Using this prescription, we have found that the first sub-leading correction to a number of quasi-amplitudes vanish. 

We would like to hight-light the following open problems:

\begin{itemize}

\item Does the protection of quasi-amplitudes in the twistor-space plane-wave sector persist to higher orders or are they a special feature of the next-to-leading order? Is there a correlation between protection and the existence of perturbatively defined transgression formulae (see discussion in Section \ref{Transgression})? We wish to stress the fact that it may in principle turn out to be the case that our regularization scheme actually yields trivial corrections, at least in the twistor-space plane-wave sector.

\item Do the quasi-amplitudes correspond to actual amplitudes of the topological open string theory in singleton phase space as proposed in \cite{Engquist:2005yt}? 
To this end, it has been proposed in \cite{Sezgin:2011hq} to complete the zero-form charges by additional contributions within a duality extended scheme that can be taken off shell such that the completions can be interpreted as the values on shell of classically marginal deformations of a topological bulk action. 

\item Turning to (possibly duality extended) quasi-amplitudes for external boundary-to-bulk states or one-body solutions \cite{Didenko:2009td,WipCarlo}, does the regularization procedure proposed here lead to a well-defined result? In that case, does it describe new sectors of free conformal field theory? For example, one may treat the standard canonical quantization of the boundary theory as a trace with an insertion of a vacuum-to-vacuum projector and then examine the effects of replacing this insertion by other operators such as for example Klein operators of the oscillator algebras of the free fields. 

\end{itemize}

Two related issues, that we also think are very interesting, concern the role our regularization procedure may have to play in the evaluation on shell of various marginal deformations in even positive degrees given by different homotopy charges defined in a soldered phase of the theory \cite{Sezgin:2011hq}: 

\begin{itemize}

\item A complex on-shell closed form of degree two is proposed in \cite{WipCarlo} to detect centers-of-mass of multi-body solutions; these solutions involve auxiliary contour integrals that can be deformed following the scheme that we have proposed.

\item An on-shell closed form of degree four is proposed in \cite{Sezgin:2011hq} as generating functional for holographic amplitudes. Upon supplementing counter terms on the boundary of spacetime and possibly also  twistor space (corresponding to a subtractive regularization scheme rather that the multiplicative one spelled out in Section \ref{Traces and observables for unbroken and metric phases}), can one apply our regularization scheme to calculate holographic correlation functions? Is it possible to relate our prescription (that deforms the auxiliary homotopy integration contours) to the prescription used successfully in \cite{Giombi:2009wh,Giombi:2010vg} (that instead deforms the contours in twistor space)?

\end{itemize}

\vspace{1cm}
\large{\bf Acknowledgements:} We have benefitted from collaborations and interactions with J. Engquist, N. Boulanger, C. Iazeolla and E. Sezgin. We also wish to thank I. Bandos, G. Barnich, X. Bekaert, A. Campoleoni, D. Chialva, V. Didenko, D. Francia, M. Grigoriev, S. Lyakhovich, K. Meissner, J. Mourad, A. Sagnotti, Ph. Spindel, F. Strocchi, M. Taronna, M. Vasiliev and Xi Yin for valuable discussions. P.~S. expresses his gratitude towards the support from Scuola Normale Superiore, Pisa during the preliminary stages of this work.

\begin{appendix}
\section{Normal-ordered symbols}
\label{normal-ordered symbols}

The perturbative expansion in twistor space is facilitated by going to the normal order that reduces to Weyl order for composites depending only on $Y^\una$ or $Z^\una$, and that preserves manifest invariance of the diagonal $\msp(4)$ subalgebra of the $\msp(4)_Y \times \msp(4)_Z$ algebra preserved by Weyl order; the basic contractions of the $\msp(4)_{\rm diag}$-invariant normal order are given by
\be \overbrace{Y_\una  Y_\unb}~=~ iC_{\underline{\a\b}}\ ,\qquad  \overbrace{Y_\una Z_\unb}~=~-iC_{\underline{\a\b}}\ ,\qquad \overbrace{Z_\una Y_\unb}~=~iC_{\underline{\a\b}}\ ,\qquad \overbrace{Z_\una Z_\unb}~=~-iC_{\underline{\a\b}}\ ,
\label{associativeYZ}
\ee
where $\overbrace{\widehat f ~\widehat g}:= \widehat f \star \widehat g- \widehat f\, \widehat g$\,. In terms of the doublets defined by $Y^\una=(y^\a,\yb^{\ad})$ and $Z^\una=(z^\a,-\zb^{\ad})$, one has
\label{associative}
\be \overbrace{y_\a y_\b}~=~i\e_{\a\b}\ ,\qquad  \overbrace{y_{\a} z_{\b}}~=~ -i\,\e_{\a\b}\ ,\qquad \overbrace{z_{\a}
y_{\b}}~=~ i\,\e_{\a\b}\ ,\qquad \overbrace{z_\a z_\b}~=~-i\e_{\a\b}\ , 
\ee
and hermitian conjugates. For polynomial polynomial composites, one thus has
 \be
 \widehat f(Y,Z)~\star~ \widehat g(Y,Z)~=~
 \int_{\cal R} \frac{d^4 S d^4 T}{(2\pi)^4}~ e^{iT^\una S_\una} ~ \widehat f(Y+S,Z+S)\widehat g(Y+T,Z-T)\ ,\label{starproductYZ}
 \ee
that is,    
\be
 \widehat f(y,\bar y;z,\bar z)~\star~ \widehat g(y,\bar
y,z;\bar z)\label{star}~=~\nonumber\ee\be
 \int_{\cal R} \frac{d^2\xi d^2\eta d^2\bar\xi
d^2\bar\eta}{(2\pi)^4}~ e^{i\eta^\a\xi_\a+
i\bar\eta^{\dot\a}\bar\xi_{\dot\a}} ~ \widehat f(y+\xi,\bar y+\bar
\xi;z+\xi,\bar z-\bar \xi)~\widehat g(y+\eta,\bar y+\bar
\eta;z-\eta,\bar z+\bar \eta)\ ,\label{starproduct}
\ee
where the $8$-dimensional auxiliary integration domain\footnote{Generic ordering schemes requires $16$-dimensional auxiliary integrals.} can be equivalently taken to be either real or chiral, \emph{viz.}
\bea\mbox{real domain $\cal R$~:}&&  (\xi_\a,\eta_\a)^\dagger~=~(\bar\xi_{\ad},\bar\eta_{\ad})\ ,\label{realdomain}\\[5pt] \mbox{chiral domain $\cal R$~:}&&  (\xi_\a,\eta_\a)^\dagger~=~(\xi_\a,\eta_\a)\ ,\qquad (\bar\xi_{\ad},\bar\eta_{\ad})^\dagger~=~(\bar\xi_{\ad},\bar\eta_{\ad})\ ,\label{chiraldomain}\eea 
giving rise to different non-polynomial extensions. In this paper we shall consider chiral extensions which are easier to implement due to the chiral nature of the Klein operators, whose normal-ordered form reads
\be 
\widehat{\kappa}~  =~ \exp(iy^{\a} z_{\a})\ ,\qquad
 \widehat{\bar{\kappa}} ~=~\widehat{\kappa} ^\dagger=\exp(-i\yb^{\ad}\zb_{\ad})\ .
\ee
In general, one always has
\bea \overbrace{Y_\una ~\widehat f\phantom{{1}}}~=~ i\left(\partial^{(Y)}_\una -\partial^{(Z)}_\una\right)\widehat f\ ,\qquad \overbrace{Z_\una~ \widehat f\phantom{{1}}}~=~ i\left(\partial^{(Y)}_\una -\partial^{(Z)}_\una\right)\widehat f\ ,\label{contractionY}\\[5pt]
\overbrace{ \widehat f ~Y_\una}~=~ -i\left(\partial^{(Y)}_\una +\partial^{(Z)}_\una\right)\widehat f\ ,\qquad \overbrace{\widehat f~Z_\una }~=~ i\left(\partial^{(Y)}_\una +\partial^{(Z)}_\una\right)\widehat f\ .\label{contractionZ}\eea

\end{appendix}

\bibliography{bibliov2}{}
\bibliographystyle{utphys}
\end{document}